# Spitzer Survey of the Large Magellanic Cloud, Surveying the Agents of a Galaxy's Evolution (SAGE) I: Overview and Initial Results

**Short Title: Spitzer Survey of the LMC, SAGE**

Margaret Meixner[1], Karl D. Gordon[2], Remy Indebetouw[3], Joseph L. Hora[4] , Barbara Whitney[5], Robert Blum [6], William Reach[7], Jean-Philippe Bernard[8], Marilyn Meade[10], Brian Babler[10], Charles W. Engelbracht[2], Bi-Qing For[2], Karl Misselt[2] , Uma Vijh[1], Claus Leitherer[1], Martin Cohen[9], Ed B. Churchwell[10], Francois Boulanger[11], Jay A. Frogel[12], Yasuo Fukui[13], Jay Gallagher[5], Varoujan Gorjian[14], Jason Harris[2], Douglas Kelly[2], Akiko Kawamura[13], SoYoung Kim[15],William B. Latter[8], Suzanne Madden[16], Ciska Markwick-Kemper[3], Akira Mizuno[13], Norikazu Mizuno[13], Jeremy Mould[17], Antonella Nota[1], M.S. Oey[18] , Knut Olsen[2], Toshikazu Onishi[13], Roberta Paladini[9], Nino Panagia[1], Pablo Perez-

[1] Space Telescope Science Institute, 3700 San Martin Drive, Baltimore, MD 21218, meixner@stsci.edu, leitherer@stsci.edu, vijh@stsci.edu, nota@stsci.edu, panagia@stsci.edu
[2] Steward Observatory, University of Arizona, 933 North Cherry Ave., Tucson, AZ 85721, kgordon@as.arizona.edu, cengelbracht@as.arizona.edu, biqing@email.arizona.edu, kmisselt@as.arizona.edu, pgperez@as.arizona.edu, jharris@as.arizona.edu, dkelly@as.arizona.edudennis@fishingholes.as.arizona.edu
[3] Department of Astronomy, University of Virginia, Astronomy Deprtment, P.O. Box 3818, Charlottesville, VA 22903-0818, remy@virginia.edu, fk2n@virginia.edu
[4] Harvard-Smithsonian, CfA, 60 Garden St., MS 65, Cambridge, MA, 02138-1516, jhora@cfa.harvard.edu
[5] Space Science Institute, 4750 Walnut St., Suite 205, Boulder, CO 80301, bwhitney@spacescience.org
[6] Cerro Tololo Interamerican Observatory, Casilla 603, la Serena, Chile, rblum@ctio.noao.edu, kolsen@ctio.noao.edu
[7] Spitzer Science Center/Jet Propulsion Lab, California Institute of Technology, 220-6, Pasadena, CA, 91125, reach@ipac.caltech.edu, vandyk@ipac.caltech.edu
[8] Centre d'Etude Spatiale des Rayonnements, Direction de la Recherche, 18 Avenue Edouard Belin, Toulouse, Cedex, F-31055, Frame, Jean-Philippe.Bernard@cesr.fr, Roberta.Paladini@cesr.fr
[9] University of California, Berkeley, Radio Astronomy Lab., 601 Campbell Hall, Berkeley, CA 94720-3411, mcohen@astro.berkeley.edu
[10] Department of Astronomy, University of Wisconsin, Madison, 475 N. Charter St., Madison, WI 53706-1582, meade@sal.wisc.edu, brian@sal.wisc.edu, ebc@astro.wisc.edu, jsg@astro.wisc.edu
[11] Astrophysique de Pari, Institute (IAP), CNRS UPR 341, 98bis, Boulevard Arago, Paris, F-75014, France, Francois.Boulanger@ias.u-psud.f
[12] AURA, Inc., 1200 New York Avenue, NW, Suite 350, Washington, DC, 20005, jfrogel@aura-astronomy.org
[13] Nagoya University, Dept. of Astrophysics, Chikusa-Ku, Nagoya, 464-01, Japan, fukui@a.phys.nagoya-u.ac.jp, kawamura@a.phys.nagoya-u.ac.jp, mizuno@a.phys.nagoya-u.ac.jp, norikazu@a.phys.nagoya-u.ac.jp, ohnishi@a.phys.nagoya-u.ac.jp, shibai@nagoya-u.jp, ssato@z.phys.nagoya-u.ac.jp
[14] JPL 480 Oak Grove Blvd., MS 169-327, Pasadena, CA 91109, varoujan.gorjian@jpl.nasa.gov
[15] Johns Hopkins University, Department of Physics and Astronomy, Homewood Campus, Baltimore, MD 21218, sykim@pha.jhu.edu
[16] Service d'Astrophysique CEA, Saclay, 91191 Gif Sur Yvette Cedex, France, smadden@cea.fr
[17] NOAO 950 N. Cherry Ave., Tucson, AZ 85726-6732, jmould@noao.edu
[18] University of Michigan, Dept. of Astronomy, 830 Dennison Bldg., Ann Arbor, MI 48109-1042 msoey@umich.edu

Gonzalez[2], Hiroshi Shibai[13], Sato Shuji[13] , Linda Smith[19], Lister Staveley-Smith[20], A.G.G.M. Tielens[21], Toshiya Ueta[22], Schuyler Van Dyk[8], Kevin Volk[23], Michael Werner[24], and Dennis Zaritsky[2]

[19] University of College, London, Dept. of Physics and Astronomy, Gower St., London, WC1E 6BT, ljs@zuaxp0.star.ucl.ac.uk

[20] Commonwealth Science and Industrial Research Organization (CSIRO), Head Office, GPO Box 4908, Melbourne, VIC, 3001 Australia, Lister.Staveley-Smith@csiro.a

[21] NASA Ames Research Center, Mail Stop 211-3, SOFIA, Moffett Field, CA 94035, tielens@astro.rug.nl

[22] URSA/NASA Ames Research Center, Mail Stop 211-3, SOFIA, Moffett Field, CA 94035, tueta@mail.sofia.usra.edu

[23] Gemini Observatory, Northern Operations Center, 670 N. A'ohuku Place, Hilo, HI, 96720, kvolk@gemini.edu

[24] JPL 480 Oak Grove Blvd., MS 264-767, Pasadena, CA 91109, mwerner@sirtfweb.jpl.nasa.gov

## Abstract

We are performing a uniform and unbiased imaging survey of the Large Magellanic Cloud (LMC, ~7°×7°), using the IRAC (3.6, 4.5, 5.8 and 8 μm) and MIPS (24, 70, and 160 μm) instruments on board the *Spitzer Space Telescope (Spitzer)* in order to survey the agents of a galaxy's evolution (SAGE), the interstellar medium (ISM) and stars in the LMC. This paper provides an overview of the SAGE legacy project including observing strategy, data processing and initial results. Three key science goals determined the coverage and depth of the survey. The detection of diffuse ISM with column densities $>1.2\times10^{21}$ H $cm^{-2}$ permits detailed studies of dust processes in the ISM. SAGE's point source sensitivity enables a complete census of newly formed stars with masses >3 $M_\odot$ that will determine the current star formation rate in the LMC. SAGE's detection of evolved stars with mass loss rates $>1\times10^{-8}$ $M_\odot$ $yr^{-1}$ will quantify the rate at which evolved stars inject mass into the ISM of the LMC. The observing strategy includes two epochs in 2005, separated by three months, that both mitigate instrumental artifacts and constrain source variability. The SAGE data are non-proprietary. The data processing includes IRAC and MIPS pipelines and a database for mining the point source catalogs, which will be released to the community in support of *Spitzer* proposal cycles 4 and 5. We present initial results on the epoch 1 data for a region near N79 and N83. The MIPS 70 and 160 μm images of the diffuse dust emission of the N79/N83 region reveal a similar distribution to the gas emissions, especially the HI 21 cm emission. The measured point source sensitivity for the epoch 1 data is consistent with expectations for the survey. The point source counts are highest for the IRAC 3.6 μm band and decrease dramatically towards longer wavelengths consistent with the fact that stars dominate the point source catalogs and that the dusty objects detected at the longer wavelengths are rare in comparison. The SAGE epoch 1 point source catalog has $\sim4\times10^6$ sources and more are anticipated when the epoch 1 and 2 data are combined. Using Milky Way (MW) templates as a guide, we adopt a simplified point source classification to identify three candidate groups, stars without dust, dusty evolved stars and young stellar objects, that offers a starting point for this work. We outline a strategy for identifying foreground MW stars, that may comprise as much as 18% of the source list, and background galaxies , that may comprise ~12% of the source list.

## 1. Introduction

The interstellar medium (ISM) plays a central role in the evolution of galaxies as the birthsite of new stars and the repository of old stellar ejecta. The formation of new stars slowly consumes the ISM, locking it up for millions to billions of years. As these stars age, the winds from low mass, asymptotic giant branch (AGB) stars and high mass, red supergiants (RSGs), and supernova explosions, inject nucleosynthetic products of stellar interiors into the ISM, progressively increasing its metallicity. This constant recycling and the associated enrichment drive the evolution of a galaxy's baryonic matter and change its emission characteristics. To understand this recycling, we have to study the physical processes of the ISM, the formation of new stars, and the injection of mass by evolved stars, and their interrelationships on a galaxy-wide scale.

Among the nearby galaxies, the Large Magellanic Cloud (LMC) is the best astrophysical laboratory for studies of the lifecycle of baryonic matter, because its proximity (~50 kpc, Feast 1999) and its favorable viewing angle (35°, van der Marel & Cioni 2001) permit studies of the resolved stellar populations and ISM clouds. The ISM in the Milky Way (MW) and in the Small Magellanic Cloud (SMC) is confused in infrared (IR) images due to crowding along the line of sight. In contrast, all LMC features are at approximately the same distance from the Sun, and there is typically only one substantial cloud along a given line of sight, so their relative masses and luminosities are directly measurable. The LMC also offers a rare glimpse into the physical processes in an environment with spatially varying sub-solar metallicity ($Z \sim 0.3$-$0.5\ Z_{\odot}$; Westerlund 1997) that is similar to the mean metallicity of the ISM during the epoch of peak star formation in the Universe (redshift of ~1.5, Madau et al. 1996; Pei et al. 1999). The dust-to-gas mass ratio has real spatial variations and is ~2-4 times lower than the value for the solar neighborhood (Gordon et al. 2003), resulting in substantially higher ambient UV fields than in the solar neighborhood. The LMC has been surveyed with many instruments revealing structures on all scales and a global asymmetry that varies with wavelength (Figure 1). The ISM gas that fuels star formation (Fukui et al. 1999; Mizuno et al. 2001; Fukui et al. 2006; Staveley-Smith 2003; Kim et al. 2003), the stellar components that trace the history of star formation (Zaritsky et al. 2004; Van Dyk et al. 1999; Nikolaev & Weinberg 2000; Holtzman et al. 1999 and Olsen et al. 1999; Harris & Zaritsky 2006), and the dust (Schwering 1989; Egan, Van Dyk & Price 2001; Zaritsky et al. 2004) have all been mapped in the LMC (Figure 1). From the perspective of galaxy evolution, the LMC is uniquely suited to study how the agents of evolution, the ISM and stars, interact as a whole in a galaxy that has undergone tidal interactions with other galaxies, the MW and SMC (Zaritsky & Harris 2004; Harris & Zaritsky 2004; Bekki & Chiba 2005).

The study of the lifecycle of galaxies has been hampered by the association of dust with the key objects driving this galactic evolution – evolved stars and protostars – and the associated extinction of the light of these objects. However, the absorbed stellar light is reradiated by the dust in the infrared (IR) and this emission provides an effective tracer of stellar mass loss, star formation, and the ISM in general. The launch of the *Spitzer Space Telescope* (*Spitzer*, Werner et al. 2004) with its sensitive detector arrays provides the necessary IR tools to survey the agents of a galaxy's evolution (SAGE), the ISM and stars, and thereby trace the lifecycle of baryonic matter. We are conducting a uniform ~7°×7° survey of the LMC in all the IRAC (3.5, 4.5, 5.8, and 8.0 μm) and MIPS (24, 70, and160 μm) bands (Figure 1). The SAGE project builds upon previous IR surveys of the LMC. The all sky IR survey by *IRAS* included 8.5°×8.5° of pointed observations of the LMC imaged at 12, 25, 60 and 100 μm (Figure 1) with an angular resolution of ~1′ and resulted in a point source list of 1823 objects (Schwering 1989). The 10°×10° MSX imaging survey of the LMC at 8.3 μm with an angular resolution of ~20″ provided a source list of 1806 objects with more precise positions than *IRAS* enabling cross-correlation with ground based near-IR surveys such as 2MASS (Egan, van Dyk & Price 2001). Both of these previous far-IR surveys revealed the most luminous dusty inhabitants of the LMC, supergiants, AGB stars, HII regions and planetary nebulae; but lacked the angular resolution and corresponding point source sensitivity to detect the more populous, less luminous sources. On the other hand, the ground based near-IR surveys of the LMC based on 2MASS at J (1.25 μm), H (1.65 μm), and $K_s$ (2.15 μm) (Nikolaev & Weinberg 2000) and DENIS at I (0.8 μm), J (1.25 μm), H (1.65 μm), and $K_s$ (2.15 μm) (Cioni et al. 2000) revealed ~820,000 and ~1.3 million sources, respectively, consisting of red giants, asymptotic giant branch stars and supergiants. These near-IR surveys detected many more sources than the far-IR surveys because of their better angular resolution and because their wavebands are more sensitive to stellar photospheres, which are more numerous than the dust-enshrouded objects. One of the purposes of SAGE is to push this IR survey work to the fainter and more numerous dusty sources in the LMC.

The remainder of this paper is organized as follows. Section 2 describes the SAGE survey including the science drivers and observing strategy. Section 3 describes the data processing approach, and the legacy aspects of SAGE. Section 4 takes a preliminary look at some of the initially processed SAGE data for a region surrounding N83 and N79, near the west end of the bar. Section 5 summarizes the paper.

## 2. SAGE Observing Program

SAGE is a uniform and unbiased ~7°×7° survey of the LMC, in all the IRAC (3.6, 4.5, 5.8 and 8 μm) and MIPS (24, 70 and 160 μm) bands using 508 total hours (291 IRAC, 217 MIPS) of the *Spitzer*. The principal characteristics of the SAGE survey are summarized in Table 1. The spatial coverage shown in Figure 1 extends beyond the infrared (IR) edge of the LMC's star formation activity and provides adequate background for calibration and measurement of non-LMC populations. The point source sensitivity estimates for the SAGE survey, listed in Table 1, improve upon previous IR surveys with MSX (Egan, van Dyk and Price 2001) and *IRAS* (Schwering 1989) by a factor of ~1000 and have better wavelength coverage. This sensitivity improvement is due to the more sensitive detectors and improved angular resolution. The angular resolution is ~2″ (0.5 pc at the distance to the LMC) in the IRAC bands, and 6″ (1.5 pc), 18″ (4.5 pc) and 40″ (10 pc) in the MIPS 24, 70 and 160 μm bands respectively. The IRAC 8 μm and MIPS 24 μm images have a 100× better areal resolution, which is proportional to angular resolution squared, than the MSX (8 μm) and *IRAS* 25 μm surveys. The MIPS 70 μm band has a 11× better areal resolution compared to the *IRAS* 60 μm survey. The MIPS 160 μm band has a 2.3× better areal resolution compared to the *IRAS* 100 μm survey. Below we describe the science drivers for the survey's characteristics and the observing strategy implemented to meet these goals.

## 2.1 Science Drivers

The SAGE survey is designed to detect the population of infrared point sources down to the confusion limit imposed by Spitzer's spatial resolution and to map, with high signal-to-noise ratio, the dust emission from diffuse and molecular clouds, photodissociation regions (PDRs), and HII regions. SAGE's coverage and sensitivity limits are driven by the science goals of the SAGE survey in three areas: the star formation, evolved stars, and ISM.

### 2.1.1 Star Formation

Star formation in the LMC appears to be a stochastic process, in which stars form in clumps, clusters and supershells (e.g. Panagia et al. 2000; Walborn et al. 1999). The star formation may be self-propagating through the energetic feedback of stellar winds and supernovae (e.g. Oey & Massey 1995; Efremov & Elmegreen 1998) but this stellar feedback also acts to eventually squelch star formation by dissipating the local ISM (Yamaguchi et al. 2001;Israel et al. 2003). The CO survey of the LMC (Fukui et al. 2001; Fukui et al. 2006) has uncovered 272 giant molecular clouds (GMCs) with similar masses and radii to MW GMCs. Current optical and near-IR observations reveal that one-third of the LMC GMCs are forming compact young massive star clusters, such as R136 in 30 Doradus, while an equally large fraction exhibit no massive star formation (Blitz et al. 2006; Fukui et al. 2006). This contrast in star formation activity may indicate a phase of deeply embedded or lower mass star formation in the latter. To date, searches for infrared young stellar objects have been targeted near current star formation. The first detected young stellar object (or protostar), N159-P1 (Gatley et al. 1982) and its surrounding has been the subject of several followups, the most recent of which is a detailed *SPITZER* study of the region by Jones et al. (2005). Studies of the Henize 206 region by Gorjian et al. (2004) and the LMC superbubble, N51D, by Chu et al. (2005) show the potential of *Spitzer* imaging in discovering and characterizing new young stellar objects in the LMC. In order to obtain an unbiased and complete census of star formation throughout the LMC, SAGE is required to be sensitive to all star formation activity from the massive star formation traced by HII regions to the lower mass star formation traced by Taurus-like complexes (Figure 2). Variability of some young stellar objects, e.g. FU Orionis systems, may be detectable in the two epochs of photometry.

### 2.1.2 Evolved Stars

High mass loss during the AGB and RSG phases leads to the formation of circumstellar envelopes that are observable via their dust emission in all IRAC and MIPS bands. Stellar mass loss can drive the late stages of stellar evolution yet the mechanism for mass loss remains poorly understood. Moreover, this mass loss is a dominant source of dust and gas return to the ISM. However, present estimates disagree on the relative contributions from these different stellar classes to the injected mass budget of a galaxy (Tielens 2001). Measuring the mass-loss rates of the entire population of evolved stars will help constrain stellar evolution modeling and the total returned mass to the LMC's ISM.

Studies based on previous infrared surveys have delved into these evolved star topics. The *IRAS* catalog was used by Loup et al. (1997) to select 198 mass-losing AGB stars. Trams et al. (1999) followed up on this *IRAS* selected sample with ISO deriving colors and chemical compositions. van Loon et al. (1999) derived mass-loss rates for these ISO sources, finding a trend of increasing mass-loss rate with luminosity. However, this *IRAS* selected sample was limited to only the most luminous AGB stars ($L > 10^4\ L_\odot$) with high mass-loss rates ($5\times10^{-6}\ M_\odot\ yr^{-1}$). Using ISOCam, Loup et al (1999) detected ~300 mass-losing AGB stars at significantly lower luminosities (~10 mag at 8 μm) and mass-loss rates, but over a limited area 0.5 square degrees in the LMC Bar.

In order to obtain a complete picture of mass loss among the LMC's evolved stars, each epoch of SAGE photometry is required to be sensitive to all evolved stars with mass loss that produces dust at significant rates; i.e. mass-loss rates $>10^{-8}\ M_\odot\ yr^{-1}$ (Figure 2). In addition, we will be able to constrain the variability of evolved stars by comparing photometry derived by analyzing the two epochs separately. The three month separation of the epochs is well-suited to detecting evolved star variability, which typically has a period of approximately one year (Wood et al. 1999).

### *2.1.3 ISM*

The dust properties in the different phases of the ISM provide insight into the evolution of the dust between phases as well as the relationship of the dust components to stellar sources of UV radiation and kinetic energy. UV extinction measurements have indicated that the dust properties in the LMC vary spatially (Gordon et al. 2003). Most of the dust mass is in the largest grains, which is traced by far-IR dust emission, such as the MIPS 70-160 $\mu$m images. Comparison of these images to the H I (Staveley-Smith et al. 2003; Kim et al. 2003) and CO data (Fukui et al. 1999), can be used to map out the dust-to-gas ratio across the LMC to search for variations. In addition to the amount of dust, the grain size distribution can be measured using the color ratios where the IRAC 3.6, 5.8 and 8 $\mu$m trace PAH emission, the MIPS 24 $\mu$m traces small grains, and the MIPS 70+160 $\mu$m traces larger grains. In particular, variations in the properties of the smallest grains, as traced by PAH emission, are of fundamental importance to the thermodynamics of the ISM because small grains are very efficient in heating the gas through the photoelectric effect (Bakes & Tielens 1994). The analysis of the *IRAS* data on the LMC indicates a lower 12 $\mu$m diffuse emission in comparison to the MW and suggests a deficit of very small dust grains, possibly due to the intense UV radiation of the LMC (Schwering 1989). SAGE's complete IRAC mapping of the lower-metallicity LMC will yield high resolution insight into recent work on the paucity of PAH emission in low metallicity galaxies (Madden 2000; Houck et al. 2004; Engelbracht et al. 2005; Galliano et al. 2005; Dale et al. 2005; Madden et al. 2006; Wu et al. 2006; O'Halloran et al. 2006). The absence of PAH and small grains will have profound influence on the gas heating and the existence of cold and warm phases in the ISM (Wolfire et al. 1995). In order to carry out these ISM studies, SAGE must be sensitive to diffuse dust emission corresponding to column densities $>1.2\text{x}10^{21}$ H $cm^{-2}$ ($A_V$=0.2 mag). Residual images, which are images with the point sources subtracted and smoothed to improve the signal-to-noise ratios, will be used for the studies of the diffuse ISM. The angular resolution achieved by SAGE is sufficient to separate the stars from the ISM and to distinguish the major cloud populations: HII regions, photodissociation regions, molecular clouds, atomic clouds and diffuse medium.

## *2.2 Observing Strategy*

### *2.2.1 Mapping Strategy*

To achieve the science and sensitivity goals, the LMC was mapped at two different epochs separated by 3 months, as detailed in Table 1. The region is many times too large to map with one Astronomical Observing Request (AOR), so the survey area was divided into smaller regions that could be efficiently planned and scheduled. For IRAC, the area was divided into 7×7 tiles of 1.1°×1.1° each, composed of 14×28 pointings of High Dynamic Range (HDR) 0.6 and 12 seconds frames with half-array steps with a total duration of 10687s per AOR. The HDR 0.6 and 12 seconds frames have corresponding exposure times of 0.4 and 10.4 seconds. We will refer to these frames as the 0.6 and 12 seconds frames throughout the paper. Mapping steps were done instead of dithers to minimize the time required to cover the desired area. This IRAC mapping technique has been used with good success on the Galactic Legacy Infrared Mid-Plane Survey Extraordinaire (GLIMPSE) *Spitzer* project (Benjamin et al. 2003). Each position in the SAGE IRAC survey has at least four frames of coverage resulting in an exposure time per pixel of at least 43.2 seconds in all IRAC bands for the complete survey and a quarter of that, 11 seconds for the single frame photometry of each epoch. The LMC was mapped in 145.5 hours per epoch, for a total of 291 hours of IRAC observing time.

For MIPS, the approximately 7.8°×7.8° region centered on the LMC was covered by 38 AORs each covering approximately 25′×4°. A MIPS AOR consists of ten 4° fast scan legs with 1/2 array cross scan steps, with a duration of 2.85 hours. The LMC is mapped with a 19x2 grid of these AORs, taking 108.5 hours per epoch, or a total of 217 hours. Tight sequential constraints relative to the roll angle rate of change have been invoked so that neighboring long strips have sufficient overlap. We have carefully designed our MIPS strategy to allow for off-source measurements in *every* scan leg which allows for accurate self-calibration of the instrumental effects. While the MIPS fast scan mode does not achieve full coverage at 160 μm, the combined epoch 1 & 2 map has a good basket weave pattern with small gaps less than a pixel in size. The well-sampled 160 μm PSF (~3 pixels per FWHM) means that interpolation can be used to fill the gaps. Each position in the SAGE MIPS survey has 20, 10 and ~3 frames of coverage at 24, 70 and 160 μm, respectively. The exposure times per pixel are 60, 30, and 6 seconds at 24, 70, and 160 μm, respectively, for the complete survey and half that for each epoch's photometry.

The mapping strategy maximizes observing efficiency while minimizing artifacts that compromise data quality and limit the scientific interpretation. The IRAC and MIPS artifacts fall in two classes: random artifacts (e.g. cosmic rays, bad pixels) and systematic artifacts that are tied to pixel location and usually systematically affect rows/columns. The random artifacts are easily removed, since our mapping strategy provides four images at each location (two overlapping images per epoch). The three-month time baseline between epochs is ideal for removal of the systematic artifacts, because it provides a 90-degree roll angle in the orientation of the detectors, which optimally removes the "striping" artifacts in MIPS and IRAC image data. In addition, these two epochs are useful constraints of source variability expected for evolved stars and some young stellar objects (YSOs).

### *2.2.2 Point Source Sensitivities*

The point source sensitivity estimates for the survey listed in Table 2 column 1 are a priori estimates derived from the *Spitzer* IRAC and MIPS exposure time calculators (ETCs), SENS-PET, for isolated point sources for a medium background. Figure 2 shows example color-magnitude diagrams for the LMC that illustrate how the sensitivity limits meet the requirements to study the two populations of greatest interest: forming stars, and evolved stars. IRAC sensitivity as faint as [8.0][25] =15$^{th}$ magnitude (0.044 mJy) allows the measurement of YSOs down to a few solar masses (the limit depends on the age, as younger YSOs of a given mass are more luminous) and of Taurus-like clusters in the LMC (Fig. 2). IRAC sensitivity as faint as [8.0] ~ 11$^{th}$ magnitude (1.7 mJy) ensures that all dusty evolved stars with mass loss rates >$10^{-8}$ $M_{\odot}$ $yr^{-1}$ are detected as inferred from ISO observations by Glass et al. (1999) and Alard et al. (2001) on the Galactic bulge, and by Ramdani & Jorissen (2002) on the low metallicity globular cluster 47 Tuc.

### *2.2.3 Surface Brightness Sensitivities*

The SAGE surface brightness sensitivity to the diffuse emission in the LMC is anticipated to be ~0.5, 1, 1, 5, and 10 MJy/sr at 5.8, 8.0, 24, 70, and 160 μm respectively with a resulting signal-to-noise ratio of ~5 per pixel based on similar observations of other nearby galaxies (e.g. M81; Gordon et al. 2004; Willner et al. 2004). The removal of artifacts is particularly important for achieving the surface brightness goal. From these diffuse emission sensitivity limits in the MIPS and IRAC 5.8 and 8 μm bands, we estimate a minimum detectable column density of ~$1.2 \times 10^{21}$ H $cm^{-2}$ ($A_V$=0.2 mag) by assuming a solar neighborhood SED for the diffuse dust emission (Desert, Boulanger & Puget 1990) and the LMC gas-to-dust ratio. The IRAC 3.6 and 4.5 μm bands also detect this same column density when their angular resolution is degraded to the 160 μm band.

[25] [ ] denotes the brightness in Vega magnitudes at the wavelength enclosed in the brackets. For example, [8.0] means the Vega magnitude at 8.0 μm.

# 3. Data Processing Approach

The full LMC mosaics of the IRAC and MIPS (Figures 3 a and b; Figure 4) data show the coverage of the SAGE survey with these two instruments over the two epochs. While the analysis of these LMC images will be the subject of future papers, we comment briefly on them here in the context of data processing. The LMC 3-color IRAC image (Figures 3a) and LMC 3-color IRAC/MIPS24 image (Figure 4) reveals the stellar component dominated by the bar at 3.6 and 4.5 μm and visually demonstrates the millions of point sources that are extracted for the SAGE catalogs. The wispy, highly sculpted dust emission of the LMC's ISM appears at 8 and 24 μm in the 3-color images (Figures 3a and 4) and at 24, 70 and 160 μm in the 3-color MIPS image (Figures 3b). The investigation of the ISM relies on well calibrated images on both small and large scales.

In the spirit of previous *Spitzer* Legacy projects, the SAGE data are non-proprietary. In addition, a uniform legacy data product, consisting of point source lists and mosaiced images, will be produced by the SAGE team for the community. In support of the *Spitzer*'s proposal Cycle 4, point source catalogs for the epoch 1 data will be made available through the *Spitzer* Science Center (SSC) well in advance of the deadline. For *Spitzer*'s proposal Cycle 5, point source catalogs and refined mosaiced images of 1°×1° tiles will be made available through the SSC. For announcement of releases, see the websites for the SSC (http://ssc.spitzer.caltech.edu/) and SAGE (http://sage.stsci.edu). The SAGE IRAC and MIPS pipelines are processing the SAGE epoch 1 and epoch 2 data separately in order to obtain the source fluxes at different times and will then stack the image data into a mosaiced image and derive final, deeper source lists from these mosaiced images. In this section, we describe the IRAC and MIPS processing for epoch 1 data.

## *3.1 IRAC Pipeline*

The Wisconsin pipeline produces two data products from the flux-calibrated IRAC data provided by the Spitzer Science Center (SSC): point source catalogs and mosaic images. The data presented in this paper were processed with the SSC pipeline version S12.4.0. The Wisconsin pipeline was originally developed to process the GLIMPSE data (Benjamin et al. 2003) and was modified for this project to handle the HDR mode. Some details can be found in the GLIMPSE pipeline documents[26] and more will be forthcoming in an IRAC pipeline processing document (in preparation).

[26]http://www.astro.wisc.edu/glimpse/docs.html

### 3.1.1 Initial Image Processing

Initial image processing steps for photometry include masking hot, dead, and missing data pixels (using SSC supplied flags). Pixels associated with saturated stars are masked using an algorithm generated by the GLIMPSE team. Several image artifacts documented by Hora et al. (2004) and the IRAC Data Handbook[27] are corrected by the Wisconsin pipeline. We correct for column pulldown, which is a reduction in intensity of the columns in which bright sources are found, in the [3.6] and [4.6] bands using an algorithm written by Lexi Moustakas (GOODS team) and modifed by GLIMPSE to handle variable backgrounds. Also in bands [3.6] and [4.5] , we use a modified version of the bright source artifact corrector[28] to improve the correction of muxbleed which is a series of bright pixels along the horizontal direction on both sides of a bright source. In the [5.8] and [8.0] bands, we correct for banding, which are streaks that appear in the rows and columns radiation away from bright sources, using an exponential function.

Two bright source artifacts are not removed: muxstripe, which is a variation in the level of column segments due to a very bright source, and latents, which is a persistence of a very bright source imaged in the prior IRAC frame.

Cosmic ray rejection routines are not applied before photometry since it often masks real stellar signal. These are applied prior to mosaic imaging, as discussed in section 3.1.6

### 3.1.2 IRAC Photometry

The epoch 1 data shown in this paper consists of two visits on each sky position for each of the long and short exposures. Photometry was performed on individual frames and then combined in the bandmerger stage. In the future, when all the data are processed, photometry will be performed on the stacked mosaic images.

We use a modified version of DAOPHOT/ALLSTAR (Stetson 1987) to perform PSF fitted point-source photometry. We iterate on the photometric calculations to improve the flux estimates. Initially, sources are found at a 3-sigma level above the local background. The local background is estimated by smoothing the image which contains sources, sky and nebular emission. The found point sources are extracted by ALLSTAR. ALLSTAR does PSF fitting, simultaneously fitting all sources found on the image. It is an iterative process, minimizing the $\chi$ value of the fit for each source. Sources that have converged and have a signal-to-noise greater than 2 are subtracted from the working image, producing a residual image. After every third iteration, the estimate of the local background for each remaining (unconverged) source is recalculated from the working image. After a maximum of 200 iterations, any remaining unconverged sources are deleted from the source list. Extracted sources then pass a second round of photometric processing. By doing small aperture photometry on the residual image at the location of every extracted source, one can assess if the extracted sources are over or under-subtracted. For benign flat sky regions, aperture photometry of the residual image produces zero flux. However in complex nebular regions where background determination is more problematic, this method has been effective in identifying sources which have been over- or under- subtracted by the ALLSTAR routine. The flux of those sources is then modified appropriately by the aperture photometry result.

[27] IRAC handbook: http://ssc.spitzer.caltech.edu/irac/dh/
[28] http://spider.ipac.caltech.edu/staff/carey/irac_artifacts

Bright cosmic rays are removed from the point source list based on an algorithm that operates on the residual images. A bright cosmic ray when extracted as a source will generally produce a characteristic signature in the residual image: a strong central peak surrounded by a negative (donut) background. Our software detects these signatures and removes the corresponding objects from the source list. Diagonal and faint cosmic rays are more problematic. Thus, our current source lists have strict S/N requirements to eliminate the faint cosmic rays from being accidentally included in our catalog, as described in section 3.1.4 .

### 3.1.3 Bandmerging to Produce Sourcelists

The point source lists are merged at two stages using a modified version of the SSC bandmerger[29]. Before the first stage, source detections with S/N less than 3 are culled. In addition sources with 0.6 second exposure time fainter than magnitude 12, 11, 9, 9 in the four IRAC bands [3.6], [4.5], [5.8] and [8.0], respectively, are culled, to prevent Malmquist bias from affecting the results. During the first stage, or in-band merge, all detections at a single wavelength are combined using position, signal-to-noise (S/N) and flux to match the sources. The 0.6 second flux is included if the signal-to-noise is greater than (5, 5, 5, 7), for the four IRAC bands [3.6], [4.5], [5.8] and [8.0], respectively. The 12 second flux is included if the magnitude is fainter than the saturation limit of (9.5, 9.0, 6.5, 6.5) for the four IRAC bands [3.6], [4.5], [5.8] and [8.0], respectively. When both criteria are met, the 0.6 and 12 second fluxes are combined, weighted by the propagated errors.

The second stage, or cross-band merge, combines all wavelengths for a given source position using only position as a criterion in order to avoid source color effects. In the cross-band merge stage, we also merge with the 2MASS catalog[30].

### *3.1.4 IRAC Catalog Criteria*

To ensure high reliability of the final point-source catalogs by minimizing the number of false sources, we adopt the following selection criteria: Given M detections out of N possible observations, we require that M/N >= 0.6 in one band, and M/N >= 0.4 in an adjacent band, with a S/N > 5, 5, 5, 7 for IRAC bands [3.6], [4.5], [5.8] and [8.0], respectively. As an example, a source is typically observed twice at 0.6 second and twice at 12 second for a total of four possible observations in each band. Such a source detected three times in band [3.6] with S/N > 5, and twice in band [4.5] with S/N > 5 would be included in the catalog.

[29] http://ssc.spitzer.caltech.edu/postbcd/bandmerge.html

[30] 2MASS: The Two Micron All Sky Survey is a joint project of the University of Massachusetts and the Infrared Processing and Analysis Center/California Institute of Technology, funded by the National Aeronautics and Space Administration and the National Science Foundation.

A source may be reliably detected in one band (usually [3.6] or [4.5]) but have questionable flux in another (usually [5.8] or [8.0]). To ensure high quality fluxes for each source, a flux/magnitude entry for a band in the catalog will be nulled, i.e. removed, for any of the four following reasons. One, the source is brighter than the saturation magnitude limits, 6.0, 5.5, 3.0, 3.0, for IRAC bands [3.6], [4.5], [5.8] and [8.0], respectively. Two, the source location is flagged as near a frame edge, in a saturated star wing or coincident with a bad pixel. Three, the S/N is less than [6, 6, 6, 10] for IRAC bands [3.6], [4.5], [5.8] and [8.0], respectively in order to mitigate Malmquist bias. Four, for 12-second only data, if M/N is less than 0.6 in order to mitigate faint cosmic ray detections. For example, if N is two (12-second only data), then M has to be equal to two or the flux is nulled. If all fluxes for a source are nulled, the source is removed from the catalog.

Note that these catalog criteria refer only to the epoch 1 catalog presented in this paper and the first publicly released data set, which is based on single-frame photometry combined during the bandmerge stage. Future catalogs, combining both epochs and based on mosaic photometry will have different criteria. Characteristics of the epoch 1 catalog are summarized in Table 2.

### *3.1.5 Absolute Photometric Calibration*

To assure that our photometric calibration is uniform across the large area observed by SAGE, and between different AORs, epochs, and wavelengths, we extract the photometry for a network of absolute stellar calibrators custom-built for SAGE. These are A0-A5V or K0-M2III stars selected from SIMBAD and their surface density within the SAGE area is approximately 1 star per 0.5 square degrees (Figure 5). The techniques used to produce the complete UV to mid-IR absolute spectra are described by Cohen et al. (2003). Any stars showing inconsistency between optical photometry, spectral type and reddening, and 2MASS photometry were culled to produce the final list of 139 viable calibrators for the epoch 1 data. The dynamic range for any IRAC band and for the MIPS 24 μm band was roughly 1000, and the faintest calibrators have magnitudes of 10.7 in IRAC and 7.7 at 24-μm. The zero points are 1594., 1024., 666.7, 277.5, 179.5, 116.5, 63.13 Jy for the magnitudes in the 2MASS J, H, Ks and IRAC bands [3.6], [4.5], [5.8] and [8.0], respectively. This calibration network goes down to 0.150 magnitude in "mean absolute deviation" as defined in Cohen et al. (2003). The entire list of calibrator stars used for verification of the SAGE catalog are listed in the electronic Table 3, a stub of which is shown in this paper, with the 2MASS and IRAC magnitudes derived for these stars.

Figure 5 shows the excellent agreement between the SAGE magnitudes and the predicted magnitudes of the calibration stars for the IRAC data. The magnitudes agree to within the one-sigma errors of our photometry or approximately 5%.

### 3.1.6 Mosaiced Images

The individual IRAC epoch 1 images are further processed prior to mosaicing. Stray light areas were not masked in this dataset, but will be masked when both epochs are included. Cosmic rays were removed using the dual-outlier rejection algorithms in the SSC Mopex package. The 12-second images were mosaiced using the MONTAGE package[31]. The resulting full LMC IRAC images for epoch 1 are shown in Figures 3a and 4.

## 3.2 MIPS Pipeline

The MIPS Data Analysis Tool v3.02 (DAT; Gordon et al. 2005) was used to do the processing and mosaicing of the individual images. The standard DAT reduction steps for MIPS 24 μm data are read2 correction, droop correction, dark subtraction, electronic nonlinearity correction, flat fielding, flux calibration, distortion correction, cosmic ray rejection, and mosaicing. The standard DAT reduction steps for the MIPS 70 and 160 μm data are read rejection (autoreject, missing, and saturated reads), electronic nonlinearity correction, ramp jump detection (cosmic rays), ramp line fitting, correction for response variations using the frequent stimulator measurements, dark subtraction, correction of the illumination pattern of the simulator, flux calibration, distortion correction, residual cosmic ray detection, and mosaicing. Details of the standard DAT reduction steps can be found in Gordon et al. (2005).

In addition to the standard reductions and before mosaicing, extra processing steps on each image were carried out using programs written specifically to improve the reduction of large, well-resolved galaxies. In particular, at 24 μm, several extra steps were included. First, possible readout offset is corrected because 1 out of 4 readouts drifts slightly. Second, the array averaged background flux value is subtracted from each MIPS AOR using data in the AOR. This "background" subtraction is intended to remove non-LMC diffuse emission, e.g. zodiacal or MW foreground diffuse emission from the fast scan leg. The background equals the average of the pixels in the MIPS AOR that are located off the LMC, as defined as the most infrared bright edge (see e.g. Figure 1, IRAS image). The subtraction makes the non-LMC emission equal to zero off of the LMC. Third, the processing excludes images affected by saturation sources, e.g. persistence that appears in a frame after a source has saturated the detector. Finally, the first five images in each scan leg are excluded because of transients associated with the boost frame. At 70 and 160 μm, a pixel dependent background was subtracted from each fast scan leg. This 70 and 160 μm "background" is composed of real emission from the zodical or MW foreground and instrumental drifts during the fast scan leg. The background is derived using a low order polynomial fit to the data in all the legs of each AOR that is outside of the LMC. At 160 μm, a spatial detection of cosmic rays is used to identify and remove them because there are too few frames to remove them by stacking the images.

[31] Montage: Montage is funded by the National Aeronautics and Space Administration's Earth Science Technology Office. See http://montage.ipac.caltech.edu

The mosaiced images presented in Figures 3b, 4, 7 and 9 are a combination of the epoch 1 and 2 data. The additional redundancy improves cosmic ray rejection and helps to minimize the striping artifacts.

### *3.2.1 MIPS Point Source Catalog*

The point sources in the mosaiced images of each AOR were extracted using the PSF fitting program StarFinder (Diolaiti et al. 2000) which is suited for well sampled PSF photometry in variable background regions that are found in MIPS data. The background in the case of these photometry measurements is the diffuse infrared emission from the LMC surrounding the source. The removal of this background from the photometry measurements is handled with the same iterative approach as the IRAC photometry albeit implemented with different programs. In these high background photometry iterations, the sky plus nebular emission background is subtracted and the photometry is done on the background subtracted images. On the first iteration, the background is estimated by smoothing the image and subtracting the smoothed image from the original image. The point sources were extracted from this background subtracted image. For the second and final iteration, the background is re-estimated by smoothing a point-source subtracted image and the photometry is performed again on the new background-subtracted image.

For each MIPS band, a point source list was created using a smoothed STinyTim, T = 100 K PSF with 3 sigma and 0.80 correlation cuts in detected sources. The MIPS point source catalogs are created by merging from each AOR, the point source lists produced from multiple AORs. For sources detected in multiple AORs, the fluxes are averaged. The zero points for the magnitudes reported in Table 1 are 7.25, 0.82 and 0.16 Jy for MIPS bands 24, 70, and 160 μm, respectively. Three separate MIPS catalogs, one for each of the MIPS bands, 24, 70 and 160 μm, were produced but not merged because the angular resolutions of these MIPS catalogs differ substantially.

## 3.3 SAGE Database

We are using a relational database management system to query the SAGE IRAC and MIPS point source catalogs. The database system is implemented with Microsoft SQL server 2000 which uses a structured query language (SQL) as the basis for the queries.

The SAGE IRAC and MIPS point source catalogs are ingested into the database system. The data can be queried using any of the catalog table parameters; for example, by position using cone searches, by flux or magnitude ranges by matching sources between catalogs or any combination of these searches. The query interface includes a simple interactive web form for small, simple queries and a more complex query system called, Catalog Archive Server Jobs System (CASJobs), an expert query tool developed by the Sloan Digital Sky Survey Collaboration. We have used the SAGE database system to create a merged catalog of the available epoch 1 data in the IRAC and MIPS 24 μm catalogs using a search box of 3″ and adopting the closest IRAC source as the match for each MIPS 24 μm source. We have also used the SAGE database system to make the queries for the initial results presented in this paper.

# 4. Preliminary Epoch 1 Results from a Region near N79 and N83

We have done a preliminary analysis of the epoch 1 SAGE data for a 1.62°×1.62° subregion centered at RA = 04h 47m 39.3 s and DEC= -69° 43′ 06.″7 which includes N79 and N83 (Figure 7). This N79/N83 region is located just off the western edge of the LMC bar as shown by the square box in Figure 1. The N79/N83 region includes both old and young stellar populations with a reasonable density and thus provides a scientifically useful gauge for results from SAGE epoch 1 data. Indeed, this field was chosen because it provided the first available, processed SAGE data that covered a range of stellar types. The N79 and N83 nebulae are HII regions that have received only minor attention in the literature; e.g. inclusion in chemical abundance studies (e.g. Vermeij et al. 2002). N83B has two highly-excited, compact blobs, N83B-1 and N83B-2 (Heydari-Malayeri et al. 2001), that are compact HII regions associated with dozens of newly formed massive stars, and suggest on going star formation activity in this region. The stellar populations surrounding N79 and N83 have not been subject to detailed infrared studies at *Spitzer* wavelengths.

## *4.1 Images*

The color images of the N79/N83 region (Figure 7) provide an overview for the differences in the IRAC and MIPS band emissions whereas the black and white images reveal the distribution of the individual IRAC (Figure 8) and MIPS (Figure 9) bands. In the IRAC 3 color image (3.6, 4.5 and 8 μm; Figure 7, top panel), the blue (3.6 μm) and green (4.5 μm) reveal the stars whereas the red (8 μm) shows mainly the diffuse dust emission. The 3-color IRAC 3.6 μm and 8 μm and MIPS 24 μm image (Figure 7, middle panel) shows how the stars and ISM dust emission relate. The density of stars, the blue point sources, drops off towards the edge of the LMC. The 8 μm dust emission appears filamentary because it traces the turbulent ISM of this star formation region. The warm continuum dust emission appears at 24 μm, which is more concentrated towards the centers of the HII regions, in contrast to the 8 μm PAH emission, which appears more diffuse and even surrounds the 24 μm emission. In the MIPS 3 color image (24, 70 and 160 μm; Figure 7, bottom panel), white point sources, which are detected in all three MIPS bands, appear in the N79 and N83 HII regions as the tips of warm dust columns pointing towards the exciting stars and are probably the bright rims of globules externally illuminated by the newly formed stars. Two white points, located in the west of the image, may be background galaxies. The red points, which have been detected in the very sensitive 160 μm band only, are most likely background galaxies. Blue points, which have been detected in the 24 μm band only, could be evolved stars or background galaxies.

The IRAC images of this N79/N83 region reveal the location of the stars and the dust emission from warm dust and PAHs (Figure 8). The 3.6 and 4.5 μm images appear similar to each other with stars dominating the images and diffuse emission apparent only in the N79 and N83 nebulae which are also the brightest regions in the longer *Spitzer* wavelengths. The origin of diffuse emission at 3.6 μm is likely to be bound-free continuum from the ionized gas with some contribution from the 3.3 μm PAH feature emission and very small dust grain emission (e.g. Engelbracht et al. 2006). The 4.5 μm diffuse emission is a combination of Brackett α, bound-free continuum and possibly very small dust grain emission. As we progress to the longer IRAC wavelengths, the diffuse emission becomes more prominent and the number of stars decreases. The 5.8 μm band emission is only slightly brighter than the 3.6 and 4.5 μm emission. The origin of the 5.8 μm diffuse emission is most likely very warm dust emission (T~600 K) with contributions from the 5.6 and 6.2 μm PAH feature emissions. The 8 μm band diffuse emission is substantially brighter and more extended than the shorter IRAC bands. Also, diffuse 8 μm band emission can be seen in regions well off the N79 and N83 HII complexes. The origin of the 8 μm diffuse emission is most likely the 7.7 and 8.6 μm PAH features.

The MIPS 24, 70 and 160 μm images show primarily dust emission of the region (Figure 9). The diagonal striping artifact at 70 and 160 μm has been minimized but not eliminated in these combined epoch 1 and 2 data. Point sources are detected in all the bands but with substantially decreasing number compared to IRAC images. At first glance, the MIPS images have similar structures with the main difference being the smoother appearance at longer wavelengths due to the lower angular resolution. Closer inspection shows that the more diffuse, cirrus dust emission that is located further from the star formation is relatively brighter at 160 μm compared to 24 μm. In the over-simplified case of blackbody dust grain emission, grains at approximately 120 K, 40 K and 20 K have peak blackbody emission at 24, 70 and 160 μm with the 120 K grains being the most emissive at all wavelengths. Thus in such an over simplified case, we expect the dust to be hottest near the massive star formation regions and emit brightly at 24 μm compared to the dust in the diffuse ISM which is significantly cooler but which would still emit brightly at 160 μm. Of course, dust grains have a size distribution with the very smallest grains or PAHs being susceptible to non-radiative equilibrium emission. A model comprising PAHs, very small grains and large grains has been successful at explaining the cirrus emission at 12 and 25 μm in the MW (e.g. Desert et al. 1990) and could potentially explain the presence of 8 and 24 μm *Spitzer* emission in the diffuse dust emission observed in the SW part in this region.

### *4.1.1 Comparison with ISM gas tracers*

One of the scientific goals for SAGE is to understand the mixing of dust into the ISM and how the dust properties vary across the LMC. For such studies it is useful to compare the dust emission to tracers of gas and here we take a preliminary look at this comparison for the N79/N83 region. Quantitative analysis of these data are beyond the scope of this paper; however, the visual comparison provides some qualitative insight. Figure 9 shows a comparison of the MIPS 24, 70 and 160 μm dust emission with three tracers of interstellar gas, Hα line emission from SHASSA (Gaustad et al. 2001), HI 21 cm line emission (Staveley-Smith et al. 2003; Kim et al. 2003) and CO J=1-0 line emission (Fukui et al. 2001; Fukui et al. 2006). At first glance, the dust emission revealed by MIPS bears more resemblance to the HI 21 cm line emission image than the Hα or CO. The diffuse extended emission found in the HI 21 cm line images is also seen in the MIPS 70 and 160 μm bands and suggests that the large dust grain emission is probably well mixed with the gas. In contrast to the HI, the Hα and CO emissions are highly structured and mostly confined to the massive star formation regions. In these HII regions, the dust emission detected by MIPS, especially the 24 μm band, exhibits very bright peaks that are not seen in the HI 21 cm line emission, but that coincide with the concentrations of Hα and CO emission. These bright 24 μm peaks reveal regions of very warm (~120 K) dust heated by the young, massive stars and the gas peaks indicated by the CO and Hα reveal that these hot dust regions coincide with high gas densities. Interestingly, the IRAC 8.0 μm emission, which trace PAHs, appears well correlated with the HI 21 cm gas emission but absent from the star formation regions suggesting that the PAHs are modified or destroyed in these regions (Figure 8). Thus the dust emission revealed by the combined MIPS 24, 70 and 160 μm bands and the IRAC 8.0 μm band traces all three phases of the ISM gas, however the relative intensity of these dust emission bands vary with the ISM environment.

### *4.2 Point Source Limits and Source Counts*

The point source lists are derived from the IRAC and MIPS data for a circular region centered at RA = 04h 47m 39.3 s and DEC= -69° 43′ 06.″7 with a 0.81° radius, which would define a circle inscribed in the IRAC and MIPS images of the N79/N83 region (Figure 7). These point source lists have been analyzed to determine the broad characteristics of the detection rates in this region and anticipated characteristics of the entire SAGE survey for epoch 1. Table 2 lists the characteristics for the SAGE survey based on the epoch 1 point source extractions for the N79/N83 region. Column 1 for Table 2 lists the SAGE band for which the following columns apply. If there is more than one SAGE band in this column, then the results are a Boolean "AND" for all the bands. The second and third columns of Table 2 list the detected fluxes and magnitudes of the faintest sources found in the epoch 1 catalogs and provide an estimate for the final epoch 1 catalog for SAGE. These epoch 1 limits are expected to be less sensitive than the final catalogs because they have less integration time than the complete survey and because systematic errors due to image artifacts are not completely removed.

The point source counts from the epoch 1catalog for this field are listed in Column 4 of Table 2. These counts are highest at 3.6 μm with 119333 detected and decrease dramatically towards longer wavelength with only 46 detections at 160 μm. This trend is consistent with expectations. The most numerous point sources, stars, show up best at 3.6 μm whereas at longer wavelengths their contributions diminish rapidly because the photospheric emission on the Rayleigh-Jeans tail declines rapidly at longer wavelengths. The longer *Spitzer* wavelengths have increased source crowding due to the larger beam and reduced contrast between the point sources and the relatively brighter background of the local ISM. The point sources that appear at the longer *Spitzer* wavelengths have a significant amount of dust emission associated with them such as young stellar objects, evolved stars and background galaxies. At the bottom part of Table 2, the source counts for combinations of *Spitzer* bands are presented. The largest number appears for sources with both IRAC 3.6 and 4.5 μm detections, as expected because these bands most effectively trace the stellar population of the region. Sources with detections in all four IRAC bands are down by a factor of ~20 from just 3.6 and 4.5 μm. When one requires detection in all IRAC bands and MIPS 24 μm, the source counts decrease by a factor of ~90 from all IRAC bands alone. The 1175 sources detected at 3.6, 8.0 and 24 μm are used for the source classification described in section 4.3. Clearly, the type and quantity of sources that we are detecting varies tremendously based on the color selection criteria. In particular, the cooler and more dusty sources, i.e. ones with MIPS counterparts are rare in comparison to the stellar population.

Lower limits for the total number of epoch 1 sources in the SAGE survey are listed in column 5 of Table 2. To derive these lower limits, we simply scaled the source counts from this N79/N83 region by $(7.1 \times 7.1) / \pi(0.81)^2 = 24.5$; i.e. ratio of the angular areas of the whole survey to that of the N79/N83 region. The total number of sources anticipated in the SAGE survey is >2.92 million in the 3.6 μm band for the epoch 1 point source catalogs. For IRAC, we can compare these counts to the total number of sources in the epoch 1 catalog that are listed in column 6 of Table 2. We find that the actual total number of sources in the epoch 1 catalog is larger by a factor of ~1.38 compared to the estimate. This discrepancy is not surprising because the N79/N83 region of the SAGE survey has a lower density of sources compared to the rest of the LMC (e.g. Westerlund, 1997). For the combined epoch 1+2 final catalog, we expect much higher detection rates because the source extraction will go deeper.

For IRAC, the epoch 1 sensitivities listed in Table 2 are largely limited by our catalog source selection criteria that rejects faint cosmic rays as well as real faint sources (see section 3.1.4). Thus, our current limiting sensitivities are higher than those predicted based on an exposure time calculator (SENS-PET), which assumed an isolated point source in a medium background, not a realistic case for all locations in the LMC. This limitation will be removed in future source lists because the epoch 1 and 2 data will be combined with cosmic ray rejection prior to photometry. Moreover, by stacking the images for the IRAC SAGE data, we will increase the final integration time by a factor of 4 and thereby improve the S/N by at least a factor of two or more if one considers the removal of the artifacts. Thus we anticipate our final point source catalog to reach the predicted sensitivities and source counts.

For MIPS, the epoch 1 point source catalog, which is still in progress for completion, the sensitivities reached for the 24, 70 and 160 μm bands (Table 2) are comparable to those predicted (Table 1). The final catalog which will combine epoch 1 & 2 is expected to meet the predicted sensitivities because the integration time for epoch 1 is only half or the final and thus a factor of ~1.4 (= √2) improved sensitivity is anticipated.

## *4.3 Source Classification Approach*

Detailed classification of all the detected sources is beyond the scope of this paper. In this section, we begin classification of the sources by applying some rough boundaries on the types of sources using templates defined for the MW. The Cohen (1993) MW templates, that were developed for *IRAS* data, have been adapted for the IRAC and 2MASS color classification scheme for GLIMPSE. Whitney et al. (2004) have developed YSO models for GLIMPSE. This MW classification scheme has been applied to the MSX sources of the LMC (Egan et al. 2001). These previous works include a wide range of infrared source types such as main sequence stars, red giants, O-rich AGB stars, C-rich AGB stars, OH/IR stars, dusty carbon stars, HII regions, planetary nebulae, young stellar objects with a range of luminosities within each of these categories.

In this preliminary work, we employ the MW templates, that were adapted for the GLIMPSE project, to classify the SAGE sources on a [3.6]-[8.0] vs. [8.0]-[24] color-color diagram (Figure 10). While they may not be completely appropriate for the LMC, these MW templates are at least a well understood starting point. On the [3.6]-[8.0] vs. [8.0]-[24] color-color diagram we plot these MW templates in a simplified manner, by grouping them into three broad categories: stars without dust (black asterisks in Figure 10), dusty evolved stars (black open triangles in Figure 10) and young stellar objects (black open square in Figure 10). Stars without dust include main sequence stars and red giants. Dusty evolved stars include the O-rich and C-rich AGB stars, OH/IR stars and carbon stars. Young stellar objects cover the range in mass and temperature and include HII regions and lower mass Class I-III sources.

The [3.6]-[8.0] vs. [8.0]-[24] color-color diagram provides the widest range in color for SAGE point sources that still includes enough sources for a valid classification. The 1175 point sources detected at 3.6, 8.0 and 24 μm in the N79/N83 region are plotted on this color-color diagram and classified into one of the three categories as follows. Using the MW templates as a guideline, we define some boundaries to regions of color space for the three categories.

1 Stars without dust are bounded by the lower-left box [8.0] – [24] < 2 and [3.6] – [8.0] < 0.5 and shown in blue.

2 Dusty evolved are enclosed in [3.6] – [8.0] < -1.525 *([8.0]-[24]) + 7.025 and [3.6] – [8.0] > 0.5 and shown in green.

3 Young Stellar Objects are defined by [3.6] – [8.0] > -1.525(8-24) + 7.025 and shown in red.

These boundaries are drawn as dashed lines on Figure 10. The points that lie in these different regions are considered candidates for this LMC sub-group and below when we state one of these subgroups by name, e.g. dusty evolved stars, we always mean candidates for this subgroup, e.g. candidate dusty evolved stars. A temperature line on this color-color plot shows that the temperature of the sources decreases from ~10,000 K to 400 K as we progress from the lower left to upper right of the diagram. Alternatively, the infrared dust emission from these sources increases in this progression. The young stellar objects are cooler than dusty evolved stars which appear cooler than stars without dust. The stars without dust appear to have two populations that change in color temperature, one near ~10,000 K which are most likely main sequence stars, and one near ~3000 K which are most likely red giant stars.

We plot two additional color-color diagrams for the N79/N83 region to see how our source classification holds up (Figure 11). All the points for plots are first plotted in black and then those sources that were included in the source classification color-color diagram (Figure 10) are over-plotted with their corresponding color classification. The IRAC [3.6]-[4.5] vs. [5.8]-[8.0] color-color diagram (Figure 11a) shows a similar separation of the source types with some mixing of sources near the boundary which only emphasizes that the boundaries we defined are not rigid, but guiding. We also see a similar trend of the young stellar objects residing at the cooler colors. The number of stars without dust has increased in this IRAC only plot. At even shorter wavelengths, the [3.6]-[8.0] vs. J-Ks color-color plot (Figure 11b) still shows relatively clear separation. The stars without dust appear in two populations, which were hinted at in the [3.6]-[8.0] vs. [8.0]-[24] color-color diagram. In comparison to the IRAC only color-color diagram, the numbers of dusty evolved stars and pre-main sequence stars are significantly less and the number of stars without dust are significantly more. This decrease is part of an overall trend in the source populations. The stars without dust are found primarily at the shorter wavelengths, and as we increase the wavelength coverage the population of dusty objects, i.e. dusty evolved stars and pre-main sequence stars, increase in numbers.

### *4.3.1 Color-Magnitude Diagrams (CMDs)*

One of the advantages of the LMC, is that all stellar populations are at essentially the same distance and thus we can use color-magnitude diagrams (CMDs) as an additional way to separate and identify the sources. Figure 12 shows four CMDs for the N79/N83 region. The top left panel shows an IRAC [4.5] vs. [3.6]-[4.5] CMD and reveals the multitude of sources detected by IRAC in this region. The top right panel shows an IRAC/2MASS CMD of [8.0] vs. [$K_s$]-[8.0] and shows how the addition of near infrared photometry can help separate the effects of dust from the stellar photosphere. The bottom two CMDs, [8.0] vs. [3.6]-[8.0] (left) and [8.0] vs. [8.0] –[24] (right), are repeats of Figure 2, but without all the MW templates. The horizontal lines in the last three CMDs marks the limit of [8.0] = 11 magnitudes below which we expect LMC AGB stars to have little or no dusty mass loss (van Loon et al. 1999; see their Figure 9, Mbol < -4.5, assuming a bolometric correction to the [8.0] of 3 magnitudes, and a distance modulus of 18.5). The diagonal edges to the points plotted on these CMDs are caused by the sensitivity and brightness limits of the epoch 1 point source catalog for SAGE. In particular, the faint limit of the 3.6 μm IRAC band affects the [4.5] vs. [3.6]-[4.5], and [8.0] vs. [3.6]-[8.0] CMDs; the faint limit of the $K_s$ of 2MASS affects the [8.0] vs. [$K_s$]-[8.0] CMD and the faint limit of the 24 μm MIPS band affects the [8.0] vs. [8.0] –[24] CMD. The maximum brightness limit at 3.6 μm also affects the [8.0] vs. [3.6]-[8.0] CMD.

The numbers of sources included in these CMDs is largest for the shortest wavelengths and decreases at the longer wavelengths (Figure 12). As with the color-color diagrams, all the points for plots are first plotted in black and then those sources that were included in the source classification color-color diagram (Figure 10) are over-plotted with their corresponding color classification. Common to all CMDs is a locus of point sources with color zero which includes mostly stars without dust, i.e. blue points. The brightest of these stars are most likely foreground MW stars (see section 4.4.2). Since the most limiting factor for inclusion in Figure 10 is detectability in the 24 μm MIPS band, the [8.0] vs. [8.0] –[24] CMD contains, almost entirely, sources from Figure 10. At shorter wavelengths the number of black points increases with decreasing wavelength and fills in the fainter end of the locus of points at zero color. This trend is consistent with the fact that the shorter wavelengths of IRAC and 2MASS are sensitive to stellar photospheres of non-dusty stars, where as the additional constraint of a 24 μm MIPS band detection makes it a source of interest for the scientific goals of SAGE.

Common to all the CMDs, the stars without dust and the dusty evolved stars appear to dominate the bluer and brighter magnitudes, respectively, while the young stellar objects tend to occupy the redder and fainter magnitudes in all four CMDs. The fact that we detect few of these faint young stellar object candidates when we include the $K_s$ of 2MASS demonstrates that the SAGE survey goes fainter than 2MASS. The young stellar object population is well separated from the stars without dust and dusty evolved star populations in the [8.0] vs. [3.6]-[8.0] and [8.0] vs. [8.0] –[24] CMDs. Although some are very bright and red, the bulk of the young stellar objects are faint and red.

In contrast to the young stellar object population, the stars without dust and dusty evolved stars have a fair amount of structure in the CMDs. These two populations are well separated in the [8.0] vs. [$K_s$]-[8.0] and [8.0] vs. [3.6]-[8.0] CMDs. The dusty evolved stars make 2 spurs to the brighter, redder part of the CMDs. This trend is seen most clearly in the [8.0] vs. [$K_s$]-[8.0] but is also present in the [4.5] vs. [3.6]-[4.5] and [8.0] vs. [3.6]-[8.0] CMDs. The stars without dust have a similar structure as the dusty evolved stars and appear as blue ends to these spurs that link the spurs to the column of stars at zero color. These sequences of dusty evolved stars are brighter than the [8.0] = 11 magnitudes limit above which dusty mass loss becomes important. The fact that the two spurs are separated in brightness may indicate two different popultations that are separated in mass. The brighter of the two spurs probably represents the dusty red supergiants, while the fainter spur is most likely the dusty AGB stars (see Figure 2). Indeed, a number of the stars in the brighter spur have supergiant identifications from the survey of Westerlund et al. (1981). These brighter sources also appear variously in the CCD photometry catalog of Massey (2002), the MSX mid infrared catalog of Egan et al. (2001), and the study of luminous AGB stars by van Loon et al. (2005). These brightest red stars are also only present in the HII regions as shown by the comparison of the CMDs for the whole N79/N83 region and those for a subset of this region that does not include the star formation regions (Figures 13 and 14). This association with ionized regions suggests that the stars of the brighter red spur are massive. A more detailed classification of these dusty evolved star populations will be discussed in a forthcoming paper by Blum et al. (in preparation).

### *4.4 Background galaxies and foreground Milky Way contributions*

The analysis of the SAGE survey requires an understanding of the MW foreground stellar contribution as well as the background extragalactic component. Although we will make better estimates of the non-LMC contributions to the catalog when some off-LMC fields are analyzed, we develop some tools here to guide this effort and derive some preliminary numbers for these contributions to the point source catalogs. Figures 13 and 14 show IRAC and 2MASS CMDs of the N79/N83 region and of the southwest (sw) quarter of this field. This southwest (sw) quarter contains all sources within a 0.41° radius circle centered on and encompassing the southwest quarter of the N79/N83 region; and because it lies off the star formation regions, it should contain mostly older field stars in the LMC plus contributions from the foreground MW stars and background galaxies. The southwest region has 1576 sources detected in [3.6] and [8.0] in comparison to the 7595 sources in the N79/N83 region.

### *4.4.1 Background Galaxies*

Figure 13 shows the [8.0] vs. [3.6]-[8.0] CMD for all sources (in black) that have been detected at these two wavelengths in the N79/N83 region. The cyan filled circles show just the subset of sources toward the south west quarter of the region. This subset will naturally exclude stars and YSOs from the star forming clusters, but include LMC stars, MW foreground stars and background galaxies. The LMC stars and MW foreground stars should populate the stars without dust and dusty evolved star regions of the CMD. We note that there is a large population of faint red sources in the YSO candidate region of the CMD in this off position (cyan filled circles) and here we discuss the possibility that all of these faint YSO candidates are background galaxies. Background galaxies in this region of the CMD have been a known contaminant for the *Spitzer* Legacy program C2D (Evans et al. 2003).

Fazio et al. (2004) have estimated the IRAC 8 μm galaxy counts in several northern fields. For two fields studied, Fazio et al. (2004) report approximately 900 galaxies per square degree at 8 μm for [8.0] < 14.0 magnitudes which is at our sensitivity limit (Table 2) for the present epoch 1 point source catalog. In Figure 13, we count 219 sources with [3.6]-[8] > 1 and 14.0 > [8] > 10 magnitudes, i.e. the YSO region, in the south west quarter which would correspond to 425 galaxies per square degree at 8 μm. This good agreement between the faint YSO candidates with the galaxy counts of Fazio et al. (2004) suggests that most of these faint candidate YSOs are background galaxies. This statistical estimation will not uniquely identify background vs LMC objects in young star forming regions. The "Taurus" point shown in Figure 2 overlaps with the background galaxy region. However, most of the objects with [3.6]-[8.0] > 1 and [8.0] > 10 magnitudes in the full survey region (black points) are likely to be galaxies. The percentage of such background galaxy candidates in the sw quarter of the N79/N83 region is 14% of the total (=219/1576) , scaling the number source density to the entire N79/N83 region suggests that 876 of the 7595 sources detected at 8 μm are background galaxies or 12% of the total sample.

### *4.4.2 Foreground Milky Way stars*

Using 2MASS near infrared photometry in conjunction with the IRAC colors, we can similarly exclude MW foreground stars. In the upper left panel of Figure 14, the [8.0] vs. Ks-[8.0] color-magnitude diagram is shown for all sources detected in the four IRAC bands plus J, H, and K in the 2MASS survey. The upper right panel ("sw") of Figure 14 shows the same color magnitude diagram, but only for sources in the southwest quarter of the N79/N83 region and excluding the background galaxies as described above ([3.6]-[8]>1 and [8]>10 magnitudes). Most of the galaxies are already excluded by the requirement that each source be detected in J, H, and K, and all four IRAC bands. The bulk of the objects lie along a vertical sequence near Ks-[8.0] = 0. These are a mixture of LMC giants and MW dwarfs and giants. A cursory check of the SIMBAD database for the bright objects near Ks-[8.0]=0 and with [8.0] < 7 magnitudes, shows a number of HD stars with giant or dwarf classifications.

Foreground MW stars may be further distinguished from LMC stars by considering the [3.6] vs. J-[3.6] color-magnitude diagram as plotted in the lower left panel of Figure 14 (also restricted to objects in the "sw" part of the N79/N83 field). These two colors span the largest wavelength baseline of the N79/N83 field photometry (IRAC and 2MASS) for which the stellar photosphere should be most important. The cool LMC giants and AGB stars are seen to extend to redder colors (roughly J-[3.6] > 0.75) compared to the bluer foreground objects and to fainter magnitudes (approximately [3.6] > 10 magnitudes). For brighter [3.6], MW giants are redder; a K0 giant has K-L = 0.75 (Koornneef 1983). Making cuts in color and magnitude represented by stars below and to the right of the *dashed* line in the lower left panel of Figure 14, the bulk of the "sw" stars are LMC AGB stars, 997 of the 1296 objects included in the "sw" diagram in Figure 14. The remaining stars are most likely foreground MW stars which number 293 in total, or 23% of this sw region sample, or 320 foreground MW stars per square degree in the field. Over the whole N79/N83 region, where the LMC source density is higher, the fraction of foreground MW stars is 18%.

The final panel (lower right) in Figure 14 shows the "sw" region of the survey again, but with the foreground stars from the [3.6] panel removed as described in the last paragraph. The mass losing stars on the AGB extend to redder and brighter magnitudes in this panel.

## 5. Summary

SAGE is a ~500 hour *Spitzer* MIPS and IRAC imaging survey of a ~7°×7° field of the LMC (Figures 3 a, b and 4) with characteristics summarized in Table 1. The SAGE data are non-proprietary and the SAGE team is committed to delivering point source lists and improved images to the SSC for community access in support of proposal cycles 4 and 5 of the *Spitzer*. The science drivers for the survey center on the lifecycle of baryonic matter in the LMC as traced by dust emission and, in particular focus on the ISM, star formation and evolved stars. We present initial results on the epoch 1 SAGE data for a region near N79 and N83 that provide a verification of the survey's goals and a start at interpreting the results. The measured point source sensitivity for epoch 1 data is consistent with expectations for the SAGE survey. For the epoch 1 catalogs, we find $3.94\times10^{6}$ sources detected in the IRAC 3.6 μm band. The images and point source counts show a similar trend. The stars dominate the light at the shortest wavelength of 3.6 μm and the diffuse dust emission becomes increasingly important at the longer wavelengths and dominates in the MIPS 24, 70 and 160 μm images. The dust emission revealed by the combined MIPS 24, 70 and 160 μm bands and the IRAC 8.0 μm band traces all three phases of the ISM gas, however the relative intensity of these dust emission bands vary with the ISM environment. Using MW templates as a guide, we adopt a simplified point source classification to identify three candidate groups, stars without dust, dusty evolved stars and young stellar objects, on the [3.6]-[8.0] vs. [8.0]-[24] color-color diagram. This source classification scheme holds up well when sources are displayed in other color-color and color-magnitude diagrams for the LMC and offers a good starting point for followup work. When plotted on CMDs, the stars without dust and dusty evolved stars separate into main sequence stars (with zero color), supergiants and AGB stars. We develop a strategy for identifying the contribution to the SAGE point source catalog of background galaxies, 155 per square degree or ~12% of sources in N79/N83 region, and foreground MW stars, 320 per square degree or ~18% of sources in N79/N83 region.

*Acknowledgments:* We are grateful to Bill Mahoney, Nancy Silberman and Lisa Storrie-Lombardi and the staff at the Spitzer Science Center for their support in implementing the SAGE observing campaigns on the *Spitzer*. Albeto Conti and Bernie Shiao of STScI were instrumental in setting up the SAGE database from which these results were extracted. We acknowledge useful discussions with Elena Sabbi. Zolt Levay of STScI created several of the color figures in this paper. This research has been funded by NASA/*Spitzer* grant 1275598 and NASA NAG5-12595.

## REFERENCES

Alard, C., et al. 2001, ApJ, 552, 289

Bakes. E. & Tielens, A.G.G.M.1994, ApJ, 427, 822

Bekki, K. & Chiba, M., 2005, MNRAS, 356, 680

Benjamin, R.A., et al. 2003, PASP, 115, 953

Blitz, L., Fukui, Y., Kawamura, A., Leroy, A., Mizuno, N., &

Rosolowsky, E. 2006, accepted in Protostars and Planet V (astro-ph/0602600)

Chu, Y.-H. et al. 2005, ApJ, 634, L189

Cioni, M.-R. et al. 2000 A&AS, 144, 235

Cioni, M.-R. L. & Habing, H.J. 2003, A&A, 402, 133

Cohen, M., Staveley-Smith, L., & Green, A. 2003 MNRAS, 340, 275

Cohen, M. 1993, AJ, 105, 1860

Costa, E. & Frogel J.A. 1996, AJ, 112, 2607

Dale, D.A., Bendo, G.J., Engelbracht, C.W. 2005 , ApJ, 633, 857

Desert, F.-X., Boulanger, F. & Puget, J.L. 1990, A&A, 237, 215

Diolaiti, E., Bendinelli, O., Bonaccini, D., Close, L., Currie, D., Parmeggiani, G. 2000, A&AS, 147, 335

Efremov, Y.N., Elmegreen, B.G. 1998, MNRAS, 299, 643

Egan, M.P., van Dyk, S.D., & Price, S.D. 2001, AJ, 122, 1844

Engelbracht, C. W.; Gordon, K. ,D.; Rieke, G. H.; Werner, M. W.; Dale, D. A.; Latter, W. B. 2005 ApJ, 628, L29

Engelbracht, C.W.; Kundurthy, P.; Gordon, K.D. et al. 2006, ApJL 642, L127

Evans, N.J. 2003, PASP, 115, 965

Fazio, G. G., et al. 2004, ApJS, 154, 39

Feast, M. 1999, IAU 190, 542

Fukui, Y., Kawamura, A., Minamidani, T., Mizuno, Y., Kanai, Y.,

Mizuno, N., Onishi, T., Yonekura, Y., Mizuno, A., & Ogawa, H. 2006, in preparation

Fukui, Y., Mizuno, N., Yamaguchi, R., Mizuno, A., Onishi, T. 2001,

PASJ, 53L, 41

Fukui, Y., Mizuno, N., Yamaguchi, R., Mizuno, A., Onishi, T., Ogawa,

H., Yonekura, Y., Kawamura, A., Tachihara, K., Xiao, K., et al. 1999,

PASJ, 51, 745

Mizuno, N., Yamaguchi, R., Mizuno, A., Rubio, M., Abe, R., Saito, H.,

Onishi, T., Yonekura, Y., Yamaguchi, N., Ogawa, H., Fukui, Y. 2001, PASJ, 53, 971

Galliano, F.; Madden, S. C.; Jones, A. P.; Wilson, C. D.; Bernard, J.-P. 2005, A&A, 434, 867

Gardiner, L.T. & Noguchi, M. 1996, MNRAS, 278, 191

Gaustad, J. E., McCullough, P.R., Rosing, W., Van Buren, D. 2001, PASP, 113, 1326

Glass, I.S., et al. 1999, MNRAS, 308, 127

Gordon, K.D., Clayton, G.C., Misselt, K.A., Landolt, A.U., & Wolff, M.J. 2003, ApJ, 594, 279

Gordon, K. et al. 2004, ApJS, 1511, 215

Gordon et al. 2004, SPIE, 5487, 177

Gordon et al. 2005, PASP, 117, 503

Gorjian et al. 2004, ApJS, 154, 275

Harris & Zaritsky 2006, in preparation

Harris, J., & Zaritsky, D. 2004, AJ, 127, 1531

Helou, G., Soifer, B. T., & Rowan-Robinson, M. 1985, ApJ, 289, L7

Heydari-Malayeri, M., Charmandaris, V., Deharveng, L., Rosa, M.R., Schaerer, D., Zinnecker, H. 2001, A&A, 372, 495

Holtzman, J. and the WFPC2 team, 1999, AJ, 118, 2262

Hora et al. 2004, Proc. SPIE, 5487, 77

Houck, J. R. et al. 2004 ApJS 154, 211

Israel, F.P. et al. 2003, A&A, 406, 817

Jones, T.J., Woodward, C.E., Boyer, M.L., Gehrz, R.D. & Polomski, E. 2005, ApJ, 620, 731

Kennicutt, , R.C. et al. 2003, PASP, 115, 928

Kim, S., Staveley-Smith, L., Dopita, M.A., Sault, R.J., Freeman, K.C., Lee, Y., Chu, Y-H 2003, ApJS, 148, 473

Koornneef, J. 1983, A&A, 128, 184

Lonsdale, C. et al. 2004, ApJS, 154, 54

Loup et al. 1999, IAU 191, 561

Loup, C.; Zijlstra, A. A.; Waters, L. B. F. M.; Groenewegen, M. A. T. 1997, A&AS, 125, 419

Madden, S. C. 2000,New AR,44,249

Madden, S.C., Galliano, F., Jones, A.P., Sauvage, M. 2006, A&A, 446, 877

Madau, P., Ferguson, H.C., Dickinson, M.E., Giavalisco, M., Steidel, C.C., Fruchter, A. 1996, MNRAS, 283, 1388

Massey, P. 2002, ApJS, 141, 81

Mizuno, N., Yamaguchi, R., Mizuno, A., Rubio, M., Abe, R., Saito, H.,

Onishi,T., Yonekura, Y., Yamaguchi, N., Ogawa, H., Fukui, Y. 2001, PASJ, 53, 971

Nikolaev , S. & Weinberg, M.D. 2000, ApJ, 542, 804

Oey, S. & Massey, P. 1995, 452, 210

O'Halloran et al. 2006, in press, astro-ph/0512404

Olsen, K. A. G. 1999, AJ, 117, 2244

Panagia, N., Romaniello, M., Scuderi, S., Kirshner, R.P. 2000, ApJ, 539, 197

Pei, Y. C., Fall, S. M., & Hauser, M.G. 1999, ApJ, 522, 604

Ramdani A. & Jorissen A. 2001, A&A, 372, 85

Reach, W. T. et al. 2005, PASP, 117, 978

Schwering, P.B.W. 1989 A&AS, 79, 105

Schultheis, M., Glass, I.S., & Cioni, M.-R. 2004, A&A, 427, 945

Smith, A.W., Cornett, R.H., & Hill, R.S. 1987, ApJ, 320, 609

Staveley-Smith, L., Kim, S., Calabretta, M.R., Haynes, R.F., Kesteven, M.J. 2003, MNRAS, 339, 87

Stetson, P. 1987, PASP, 99, 191.

Tielens, A.G.G.M. 2001, ASPC, 231, 92

Trams, N. R. et al. 1999, 346, 834

van der Marel, R. & Cioni, M.-R. L. 2001, AJ, 122, 1807

Van Dyk, S.D., Cutri, R., Weinberg, M.D., Nikolaev, S., Sktutskie, M.F. 1999, in IAU symp. 190, p. 363

van Loon, J.T., Groenewegen, M.A.T., de Koter, A., Trams, N.R., Waters, L.B.F.M., Zijlstra, A.A., Whitelock, P.A., & Loup, C. 1999, A&A, 351, 559

van Loon et al. 2005, A&A, 442, 597

Vermeij, R., v& an der Hulst, J.M. 2002, A&A, 391, 1081

Walborn, N.R., Barba, R.H., Brandner, W., Rubio, M., Grebel, E.K., Probst, R.G. 1999, AJ, 117, 22

Westerlund, B.E., Olander, N., & Hedin, B. 1981, A&AS, 43, 267

Westerlund, B.E., 1997, *The Magellanic Clouds*, (Cambridge University Press: New York), p. 234

Whitney, B., Indebetouw, R. Bjorkman, J.E. & Wood, K. 2004, ApJ, 617, 1177

Willner, S. P., Ashby, M. L. N.; Barmby, P. et al. 2004, ApJS, 154, 222

Wolfire, M., Hollenbach, D., McKee, C.F., Tielens, A.G.G.M., Bakes, E.L.O. 1995 ApJ, 443, 152

Wood, P.R., Alock, C., Allsman, R.A. et al. 1999, in the Proceedings of Asymptotic Giant Branch Stars, IAU Symposium #191, eds. T. Le Bertre, A. Lebre and C. Waelkens, p. 151

Wu et al. 2006, in press (astro-ph/0510856)

Yamaguchi,R. et al. 2001, PASJ, 53, 985

Zaritsky, D., Harris, J., Thompson, I.B., & Grebel, E.K. 2004, AJ, 128, 1606

Zaritsky, D. & Harris, J. 2004, ApJ , 604, 167

## Figure Captions

**Figure 1:** Examples of the many existing LMC surveys: HI (Staveley-Smith et al. 2003; Kim et al. 2003), CO (Fukui et al. 2001; Fukui et al. 2006), *IRAS* 100 μm, Hα (Gaustad et al. 2001) with the SAGE-MIPS coverage overlaid, UV (Smith, Cornett and Hill 1987) and stellar density (Zaritsky et al. 2004) with the SAGE-IRAC coverage overlaid. The SAGE survey maps the LMC twice. Epoch 1 coverage is outlined in green and epoch 2 coverage is outlined in black. The location of the N79/N83 region, discussed in section 4, is outlined by the square box in the *IRAS* 100 μm image.

**Figure 2:** These Spitzer color-magnitude diagrams show the discovery space for SAGE. [] denotes the brightness in Vega magnitudes at the wavelength enclosed in the brackets. For example, [8.0] means the Vega magnitude at 8.0 μm. The symbols identify populations of key sources throughout the LMC: YSOs (1-30 $M_{\odot}$), HII regions, Taurus-like clusters, O-rich and C-rich AGB stars, RSGs and main sequence O stars. Symbols, as noted in the legend, represent template/model photometry of Cohen (1993) and Whitney et al. (2004). SAGE's sensitivity limit (solid line) falls x1000 below the MSX limit (dashed line) and the lower limit to AGB mass loss, $>10^{-8}$ $M_{\odot}$ $yr^{-1}$ (dotted line). The (yellow) filled circle with a star represents a subregion of Taurus containing ~12 stars, placed at the distance of the LMC.

**Figure 3a:** Full LMC mosaics of the IRAC epoch 1 data of SAGE showing the location of the surveyed region on the sky. The IRAC data is shown in three colors with IRAC-1, 3.6 μm, in blue, IRAC-2, 4.5 μm, in green and IRAC-4, 8 μm, in red.

**Figure 3b:** Full LMC mosaics of the MIPS epoch 1 and 2 data of SAGE showing the location of the surveyed region on the sky. The MIPS data is shown in three colors with 160 μm in red, 70 μm in green and 24 μm in blue.

**Figure 4:** Full LMC mosaic of the SAGE data shown in three colors: epoch 1 IRAC-1 3.6 μm in blue, IRAC-4 8 μm in green, and MIPS 24 μm in red.

**Figure 5:** The locations of the SAGE absolute photometric standards calibration network listed in Table 3. The trapezoid box is the approximate location of the SAGE survey area adequate to cover SAGE's IRAC and MIPS observations. The two circles represent stars within roughly 1deg and 2deg of the south ecliptic pole. Any such stars will be of gratuitous value to other missions observing the Ecliptic poles such as WISE and JWST. Both of these are subsets of the short-dashed box, which encompasses the entire field of view, that indicates the area searched for appropriate calibration stars using SIMBAD. Symbols indicate the quality of each stellar calibrator with the best being the filled circles, followed by filled squares, then filled triangles. Those show consistency between optical and NIR (2MASS) photometry and the spectral type and extinction to within 0.010 mag, 0.030 mag., 0.060 mag. respectively. Open squares, open triangles, and open stars represent lower-ranked calibrators with consistency at the levels of 0.090, 0.12, and 0.15 mag, respectively.

**Figure 6:** Plots demonstrating the quality of the SAGE flux and magnitude measurements in the epoch 1 IRAC point source catalogs. For each IRAC band, noted in the top left of the plots, the measured SAGE magnitudes are plotted against the predicted magnitudes for the 139 calibration stars in the SAGE field. The size of the crosses represent the error bars on both axes.

**Figure 7:** Three color SAGE images of the N79/ N83 region near the SW end of the LMC Bar, *Top:* IRAC 3-color image: IRAC-1, 3.6 μm, in blue, IRAC-3, 4.5 μm, in green and IRAC-4, 8 μm, in red. *Middle:* IRAC 3.6 μm in blue, IRAC 8 μm in green and MIPS 24 μm in red. *Bottom :* MIPS 3-color image: 160 μm in red, 70 μm in green and 24 μm in blue. In all the images, white is a combination of all three colors.

**Figure 8:** The IRAC epoch 1 SAGE data of the region surrounding N79/N83: *Top left:* IRAC 3.6 μm, *Top right:* IRAC 4.5 μm, *Bottom left:* IRAC 5.8 μm and *Bottom right:* IRAC 8.0 μm.

**Figure 9:** The MIPS epoch 1 SAGE data of the N79/N83 region in comparison with ISM gas tracers. *Left column*, *Top:* MIPS 24 μm with a square-root greyscale ranging from 0 to 140 MJy/sr*, Middle:* MIPS 70 μm with square-root greyscale ranging from 0 to 130 MJy/sr *Bottom:* MIPS 160 μm with a square-root greyscale ranging from 0 to 350 MJy/sr . *Right column*, *Top:* the Hα SHASSA data (Gaustad et al. 2001), *Middle:* the HI 21 cm line data (Staveley-Smith et al. 2003; Kim et al. 2003), and *Bottom:* the CO J=1-0 line data where the contour marks the 3σ level and the data are absent from the bottom right (southwest) part of the image (Fukui et al. 2001; Fukui et al. 2006).

**Figure 10:** The [3.6]-[8.0] vs. [8.0]-[24] color-color diagram shows the broadest separation of sources while still retaining enough sources to derive a source classification. For all bands, [ ] means the Vega magnitude at that wavelength, e.g. [8.0] is the Vega magnitude at 8.0 μm. Black symbols are MW templates from Cohen et al. (1993) and Whitney et al. (2004). Asterisks are stars without dust, such as main sequence stars, or red giants. Open triangles are dusty evolved stars. Open squares are young stellar objects. The black dashed lines mark the boundaries for the source classification based on the location of the MW templates. The orange dashed line shows the color temperature for blackbodies with temperatures ranging from 10,000 K to 400 K, marked as grey crosses on the line. The 1175 colored dots are point sources detected in the IRAC 3.6, and 8.0 μm and the MIPS 24 μm from the N79/N83 region. The red, green and blue dots are candidate young stellar objects, dusty evolved stars and stars without dust, respectively. These color classified sources are plotted in the color-color and CMDs of Figure 9 and 10, respectively, to provide guidance on the types of sources plotted.

**Figure 11:** Color-color plots, *Left:* the IRAC only and *Right:* the IRAC and 2MASS, showing how the source classification holds at other wavelengths for point sources

detected in the SAGE N79/N83 region. The black dots are plotted first and include all the sources detected in bands plotted. The colored dots are overlaid on the black dots and include the same dots as plotted in the source classification template (Figure 8).

**Figure 12:** Four CMDs of SAGE data showing all the point sources from the N79/N83 region in black which have been detected at the wavelengths plotted in the CMD. The candidate classified sources from the source classification are overlaid in the same color used in Figure 8. The black points are sources detected in the bands of the CMD but not in all of the bands of the source classification diagram and hence tend to be fainter sources not detected at the longer wavelengths, e.g. 24 μm. The black dashed line at [8.0] = 11 magnitudes marks the mass-loss rate of $10^{-8}$ $M_{\odot}$ $yr^{-1}$, above which the AGB stars appear to have dusty winds.

**Figure 13:** The [8] vs. [3.6]-[8] CMD for all sources detected in the N79/N83 region at these two wavelengths. Overplotted in cyan filled circles is the subset of sources in the south west quarter of the N79/N83 region that is outside of the HII regions.

**Figure 14:** Near infrared (2MASS) and IRAC CMDs for sources detected in all seven bands in the N79/N83 region (*upper left* panel). The "sw" region (*upper right* panel) contains sources detected in the N79/N83 southwest quarter and has been corrected for background galaxies based on the cuts of Figure 11, see text. The [3.6] vs. J-[3.6] diagram (*lower left* panel) is effective in discriminating between foreground MW objects and LMC giants and AGB stars which should generally lie below and to the right of the *dashed* line (see text). The "corrected sw" diagram shows the remaining sources, primarily LMC AGB stars, after removing foreground MW stars and background galaxy contributions.

# Tables

**Table 1: Principal Characteristics for SAGE Survey, *Spitzer* program ID 20203**

| Characteristic | IRAC Value | MIPS Value |
|---|---|---|
| Nominal Center point | RA(2000): 5h 18m 48s<br>Dec(2000): -68° 34′ 12″ | RA(2000): 5h 18m 48s<br>Dec(2000): -68° 34′ 12″ |
| survey area | 7.1°×7.1° | 7.8°×7.8° |
| AOR size, grid size | 1.1°×1.1°, 7×7 | 25′×4°, 19×2 |
| Total time (hrs) | 290.65 | 216.84 |
| λ (μm) | 3.6, 4.5, 5.8 and 8 | 24, 70 and 160 |
| pixel size at λ | 1.2″, 1.2″, 1.2″,1.2″ | 2.5″, 9.8″, 15.9″ |
| angular resolution at λ | 1.7″, 1.7″, 1.9″, 2″ | 6″, 18″, 40″ |
| Exposure time/ pixel at λ (s) | 43, 43, 43, 43 | 60, 30, 6 |
| Predicted point source sensitivity, 5 σ at λ (mJy) | 0.0051, 0.0072, 0.041, 0.044 | 0.5, 30, 275 |
| Predicted point source sensitivity, 5 σ at λ (mag.) | 19.3, 18.5, 16.1, 15.4 | 10.4, 3.5, -0.6 |
| Saturation limits (Jy) at λ | 1.1, 1.1, 7.4, 4.0 | 4.1, 23, 3 |
| Saturation limits (mag) at λ | 6, 5.5, 3.0, 3.0 | 0.60, -3.7, -3.2 |
| Surface brightness limits (MJy/sr) 5 σ at λ (mag.) | … , … , 0.5, 1 | 1, 5, 10 |
| Epoch 1 | July 15 – 26, 2005 | July 27 – Aug. 3, 2005 |
| Epoch 2 | Oct. 26 – Nov. 2, 2005 | Nov. 2-9, 2005 |

**Table 2: Epoch 1: Initial Results Point source Sensitivities and Source counts**

| λ (μm) | Epoch 1 Limiting flux [mJy] | Epoch 1 Limiting Magnitude | # sources detected in N83/N79 field | Lower limit to # sources predicted in the epoch 1 SAGE survey | # of sources in the SAGE IRAC epoch 1 point source catalog |
|---|---|---|---|---|---|
| 3.6 | 0.0127[a] | 18.348 | 119333 | $2.92\times10^{6}$ | $3.94\times10^{6}$ |
| 4.5 | 0.01839[b] | 17.474 | 59129 | $1.45\times10^{6}$ | $2.00\times10^{6}$ |
| 5.8 | 0.1001[c] | 15.164 | 13292 | $3.26\times10^{5}$ | $4.61\times10^{5}$ |
| 8 | 0.1288[d] | 14.226 | 7763 | $1.90\times10^{5}$ | $2.60\times10^{5}$ |
| 24 | 0.211[e] | 11.34 | 5115 | $1.25\times10^{5}$ | … |
| 70 | 23.9[f] | 3.8 | 1024 | $2.50\times10^{4}$ | … |
| 160 | 142[g] | 0.12 | 46 | 1130 | … |
| 3.6,4.5 | … | … | 56599 | $1.39\times10^{6}$ | $1.90\times10^{6}$ |
| All IRAC | … | … | 6737 | $1.65\times10^{5}$ | $2.26\times10^{5}$ |
| IRAC+2MASS J,H,K | … | … | 6422 | $1.57\times10^{5}$ | $2.15\times10^{5}$ |
| IRAC+24 | … | … | 627 | $1.57\times10^{4}$ | … |
| 3.6, 8.0 24 | … | … | 1175 | $2.88\times10^{4}$ | … |

[a] 6 σ, [b] 6 σ, [c] 6 σ, [d] 10 σ, [e] 3 σ, [f] 3 σ, [g] 3 σ

**Table 3: Caption: SAGE Calibration Stars for the IRAC photometry, RA and DEC in decimal degrees, electronic version has the entire list. This table is not included in this paper. See AJ paper when published for this electronic table.**

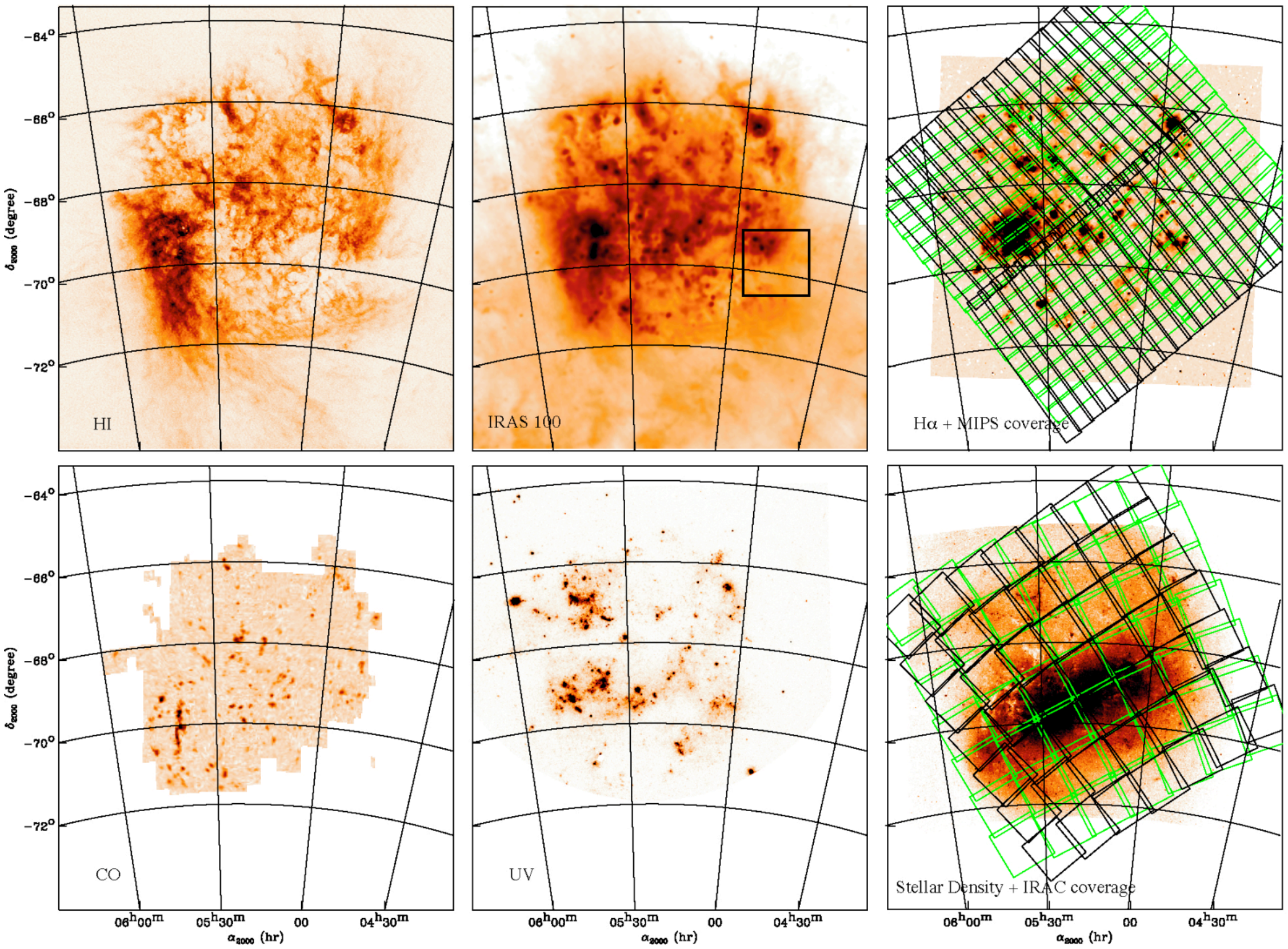

**Figure 1** Examples of the many existing LMC surveys: HI (Staveley-Smith et al. 2003; Kim et al. 2003), CO (Fukui et al. 2001; Fukui et al. 2006), *IRAS* 100 μm, Hα (Gaustad et al. 2001) with the SAGE-MIPS coverage overlaid, UV (Smith, Cornett and Hill 1987) and stellar density (Zaritsky et al. 2004) with the SAGE-IRAC coverage overlaid. The SAGE survey maps the LMC twice. Epoch 1 coverage is outlined in green and epoch 2 coverage is outlined in black. The location of the N79/N83 region, discussed in section 4, is outlined by the square box in the *IRAS* 100 μm image.

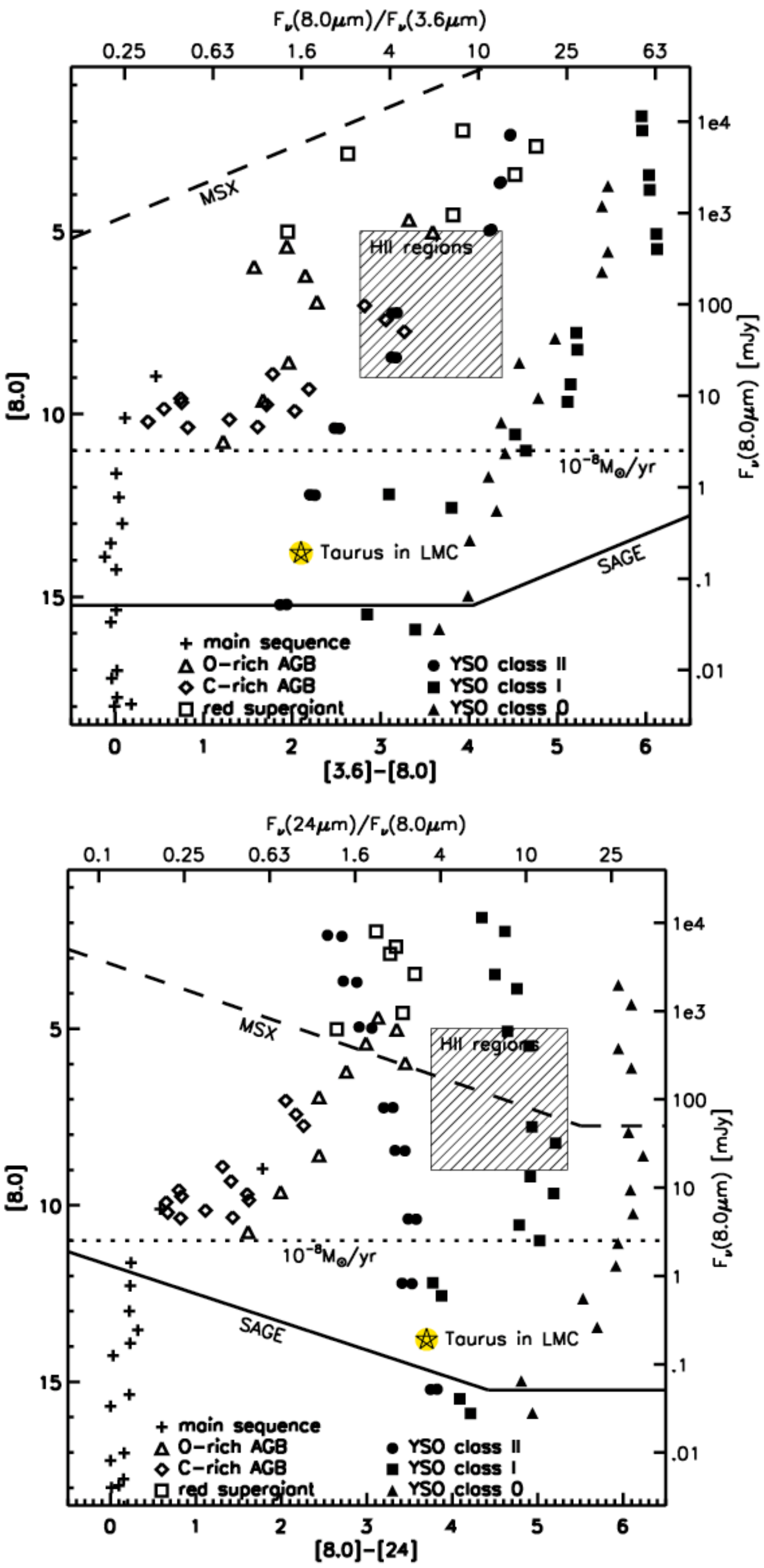

**Figure 2** These Spitzer color-magnitude diagrams show the discovery space for SAGE. [] denotes the brightness in Vega magnitudes at the wavelength enclosed in the brackets. For example, [8.0] means the Vega magnitude at 8.0 μm. The symbols identify populations of key sources throughout the LMC: YSOs (1-30 $M_\odot$), HII regions, Taurus-like clusters, O-rich and C-rich AGB stars, RSGs and main sequence O stars. Symbols, as noted in the legend, represent template/model photometry of Cohen (1993) and Whitney et al. (2004). SAGE's sensitivity limit (solid line) falls x1000 below the MSX limit (dashed line) and the lower limit to AGB mass loss, >$10^{-8}$ $M_\odot$ $yr^{-1}$ (dotted line). The (yellow) filled circle with a star represents a subregion of Taurus containing ~12 stars, placed at the distance of the LMC.

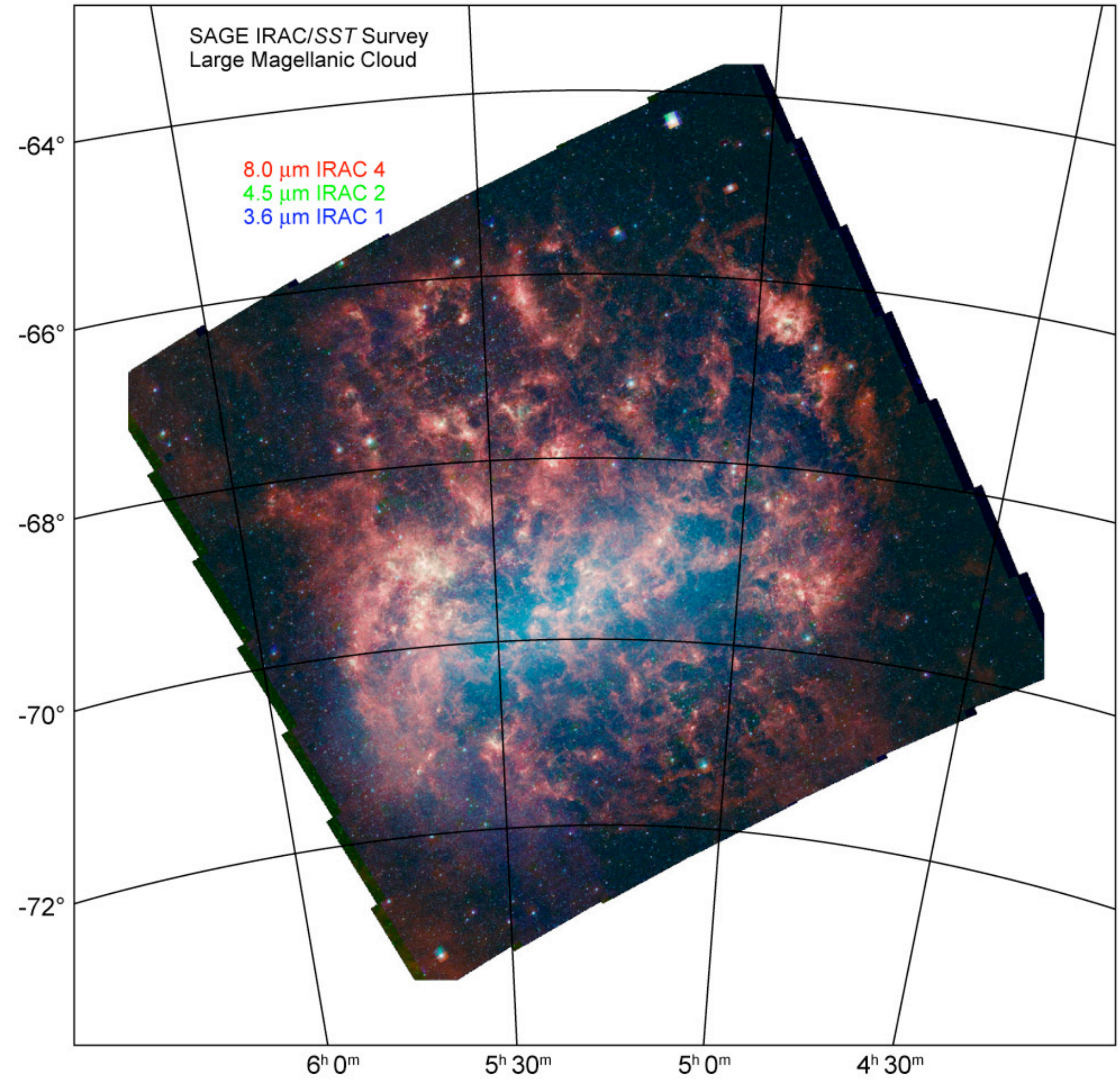

**Figure 3 a** Full LMC mosaics of the IRAC epoch 1 data of SAGE showing the location of the surveyed region on the sky. The IRAC data is shown in three colors with IRAC-1, 3.6 μm, in blue, IRAC-2, 4.5 μm, in green and IRAC-4, 8 μm, in red.

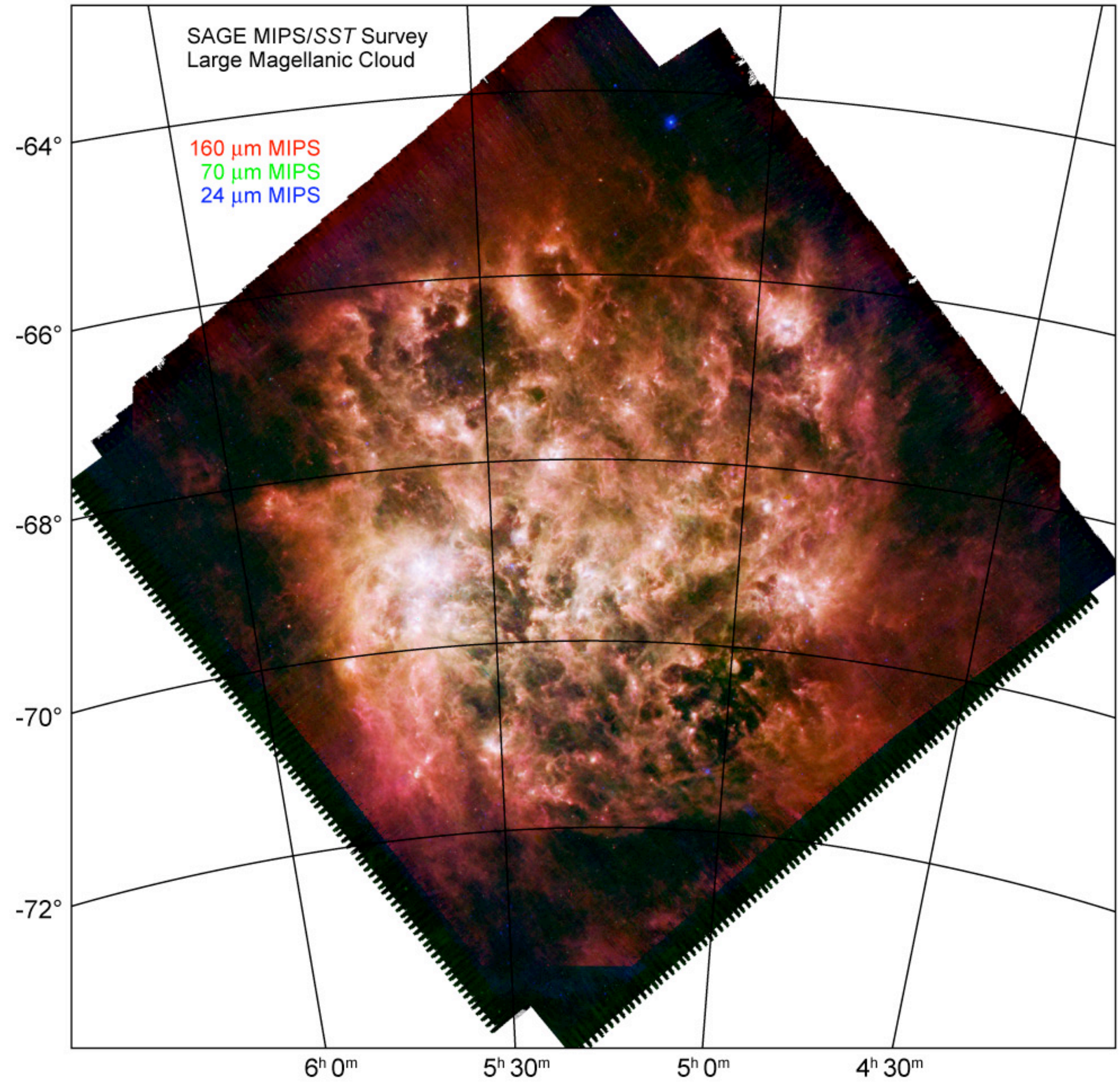

**Figure 3b** Full LMC mosaics of the MIPS epoch 1 and 2 data of SAGE showing the location of the surveyed region on the sky. The MIPS data is shown in three colors with 160 μm in red, 70 μm in green and 24 μm in blue.

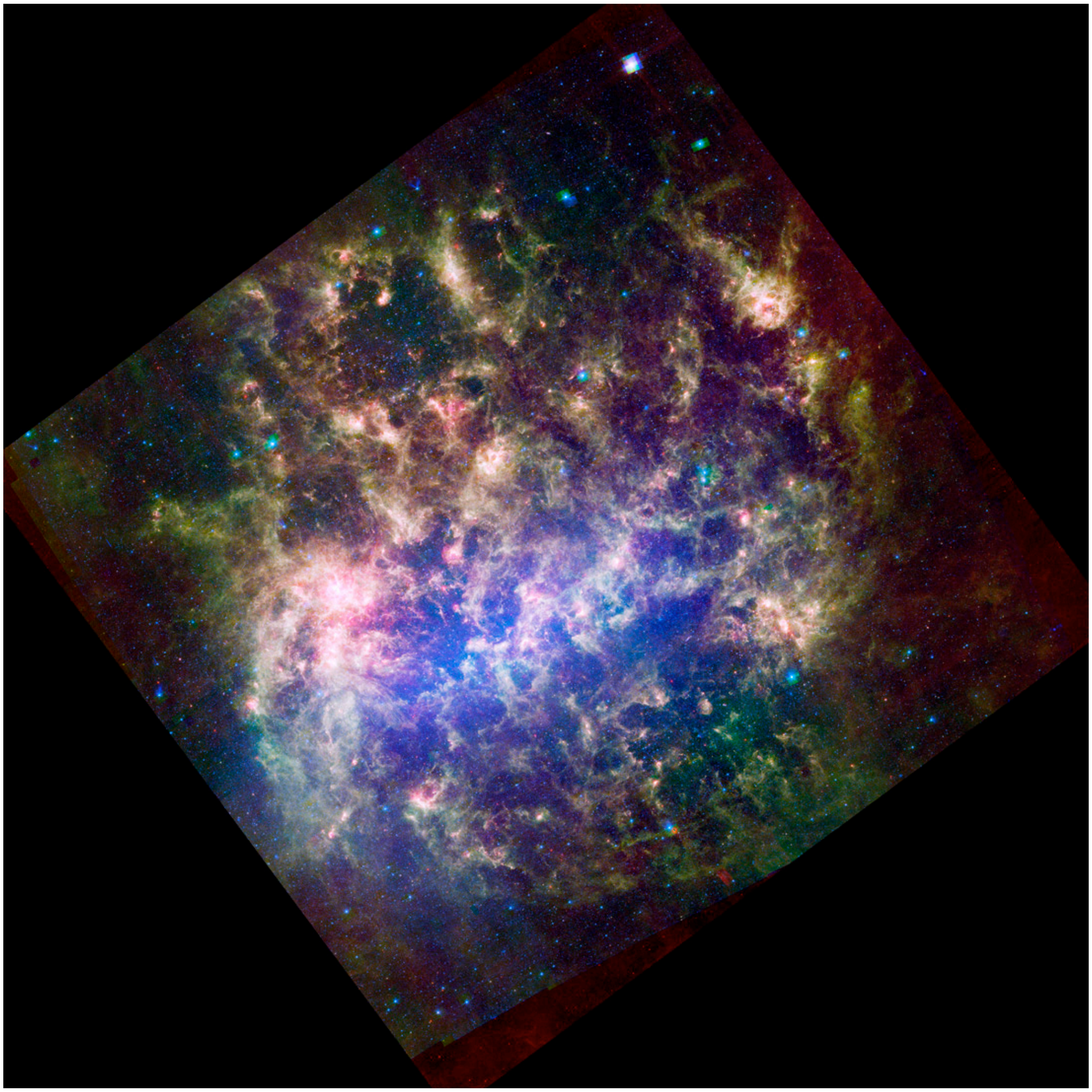

**Figure 4:** Full LMC mosaic of the SAGE data shown in three colors: epoch 1 IRAC-1 3.6 μm in blue, IRAC-4 8 μm in green, and MIPS 24 μm in red.

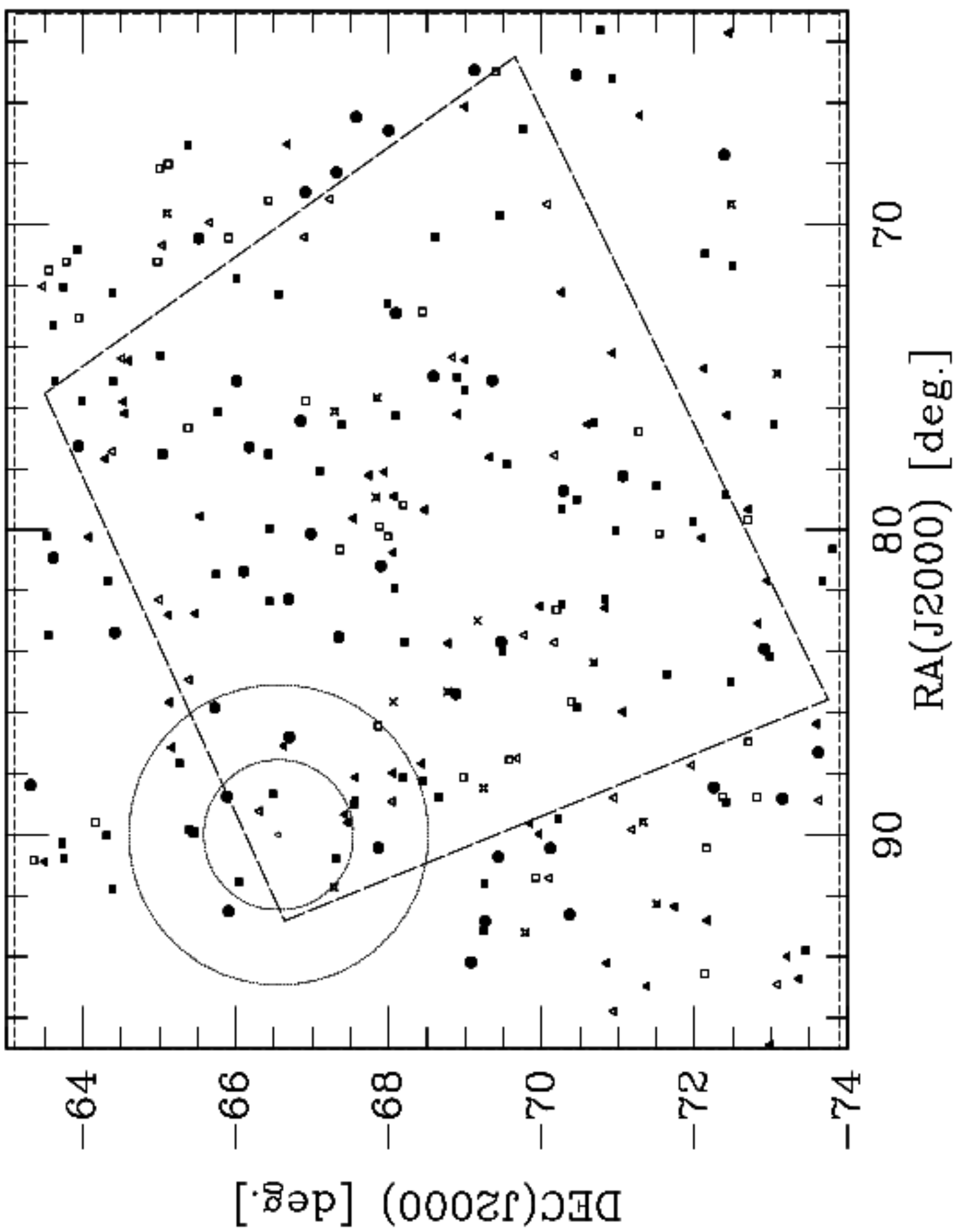

**Figure 5** The locations of the SAGE absolute photometric standards calibration network. The trapezoid box is the approximate location of the SAGE survey area adequate to cover SAGE's IRAC and MIPS observations. The two circles represent stars within roughly 1deg and 2deg of the south ecliptic pole. Any such stars will be of gratuitous value to other missions observing the Ecliptic poles such as WISE and JWST. Both of these are subsets of the short-dashed box, which encompasses the entire field of view, that indicates the area searched for appropriate calibration stars using SIMBAD. Symbols indicate the quality of each stellar calibrator with the best being the filled circles, followed by filled squares, then filled triangles. Those show consistency between optical and NIR (2MASS) photometry and the spectral type and extinction to within 0.010 mag, 0.030 mag., 0.060 mag. respectively. Open squares, open triangles, and open stars represent lower-ranked calibrators with consistency at the levels of 0.090, 0.12, and 0.15 mag, respectively.

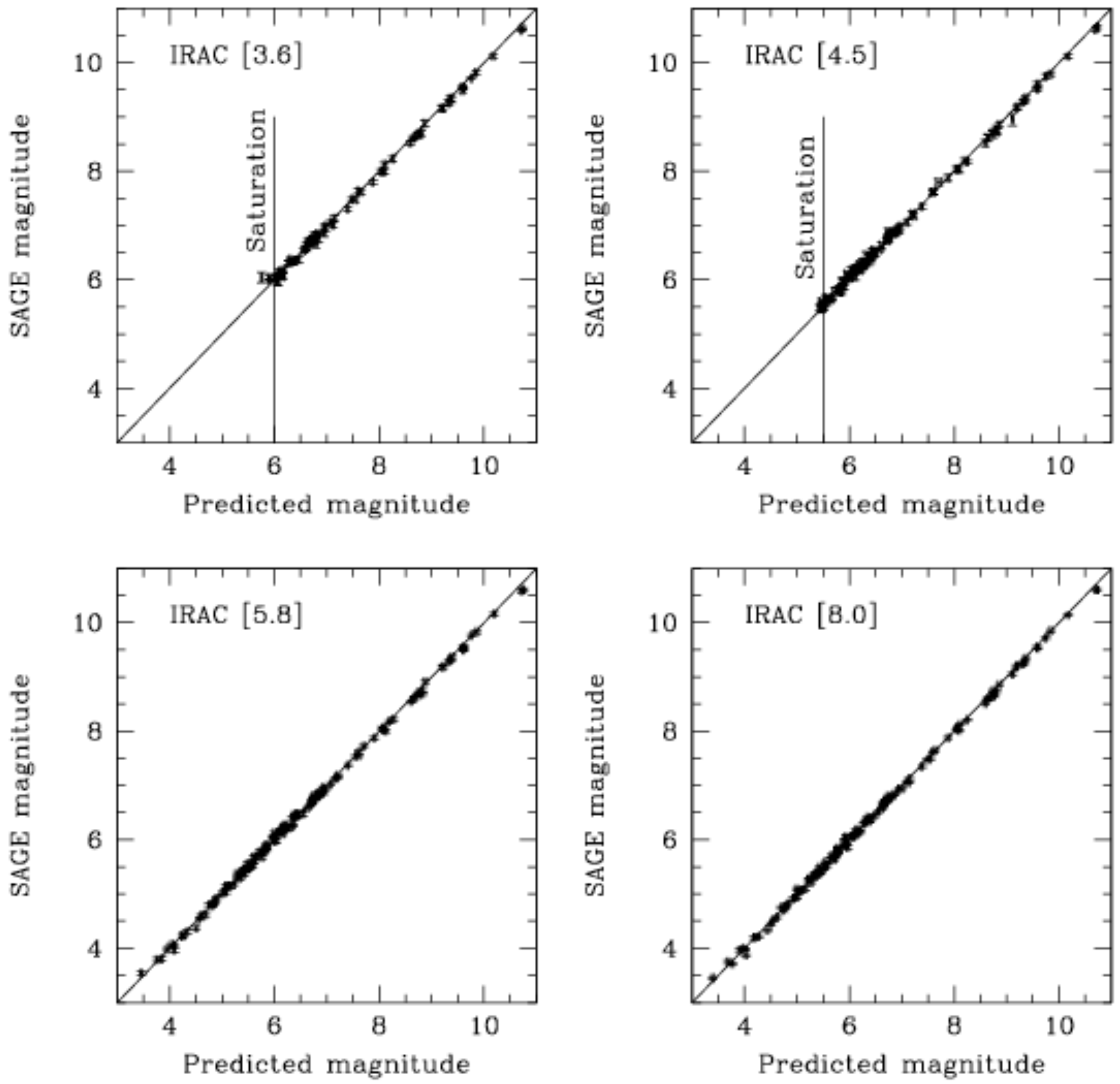

**Figure 6** Plots demonstrating the quality of the SAGE flux and magnitude measurements in the epoch 1 IRAC point source catalogs. For each IRAC band, noted in the top left of the plots, the measured SAGE magnitudes are plotted against the predicted magnitudes for the 139 calibration stars in the SAGE field. The size of the crosses represent the error bars on both axes.

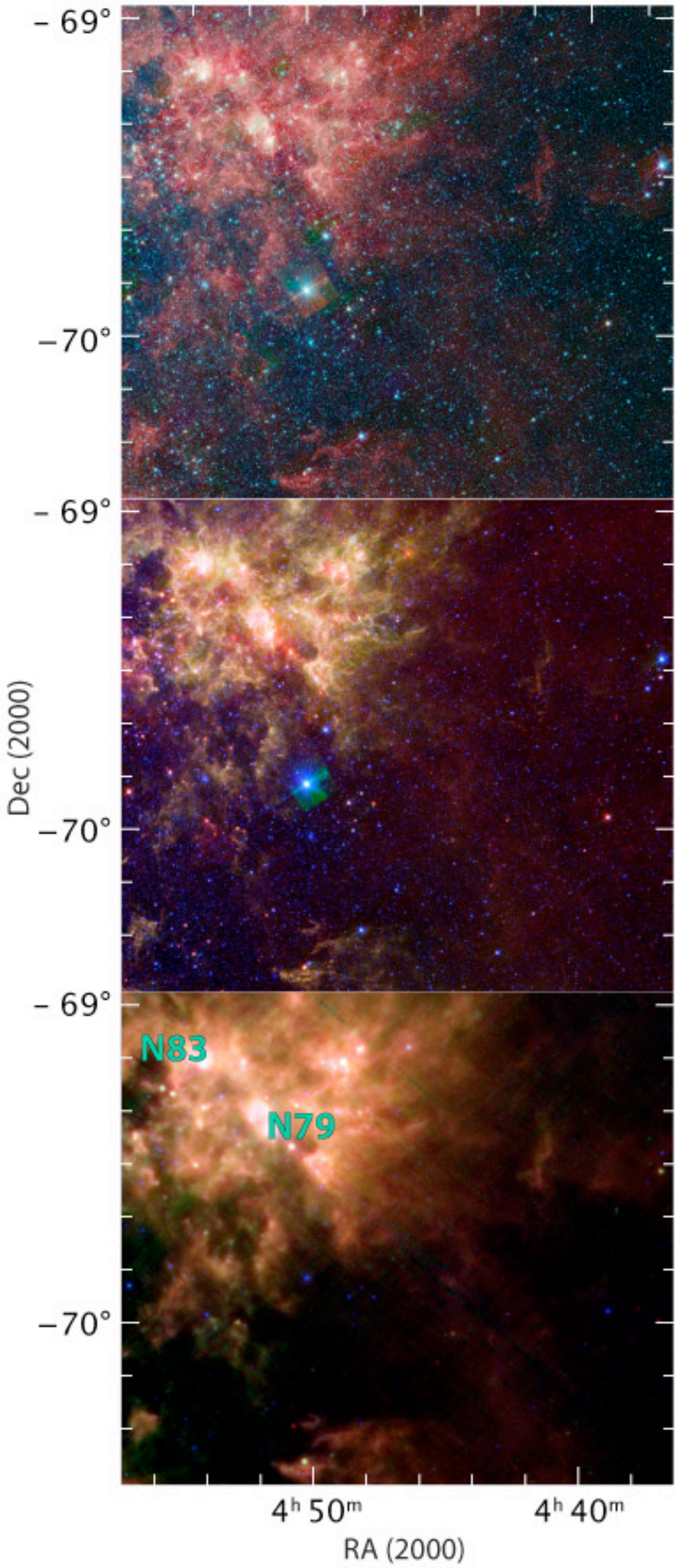

**Figure 7** Three color SAGE images of the N79/ N83 region near the SW end of the LMC Bar, *Top:* IRAC 3-color image: IRAC-1, 3.6 μm, in blue, IRAC-3, 4.5 μm, in green and IRAC-4, 8 μm, in red. *Middle:* IRAC 3.6 μm in blue, IRAC 8 μm in green and MIPS 24 μm in red. *Bottom :* MIPS 3-color image: 160 μm in red, 70 μm in green and 24 μm in blue. In all the images, white is a combination of all three colors.

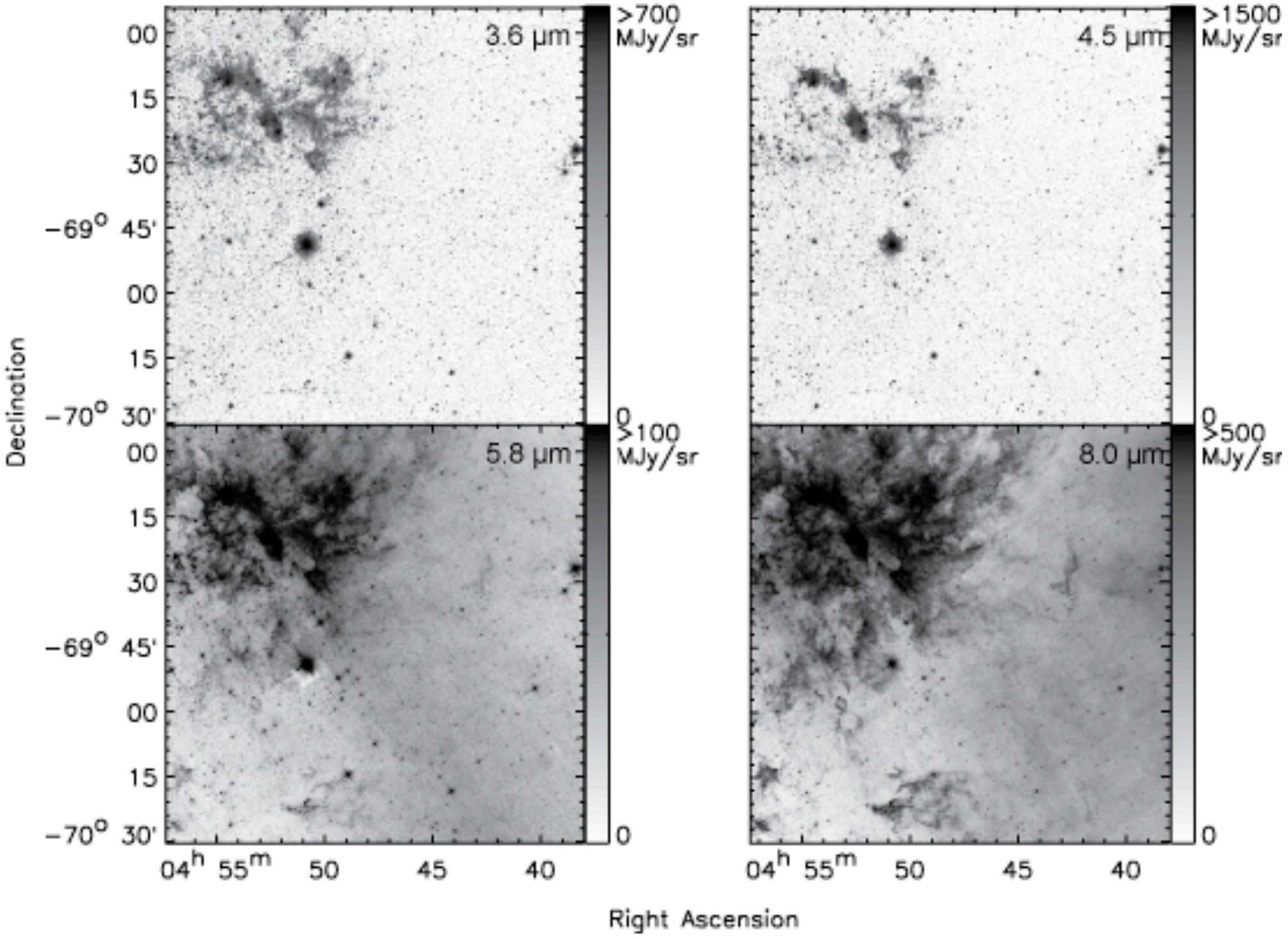

**Figure 8** The IRAC epoch 1 SAGE data of the region surrounding N79/N83: *Top left:* IRAC 3.6 µm, *Top right:* IRAC 4.5 µm, *Bottom left:* IRAC 5.8 µm and *Bottom right:* IRAC 8.0 µm.

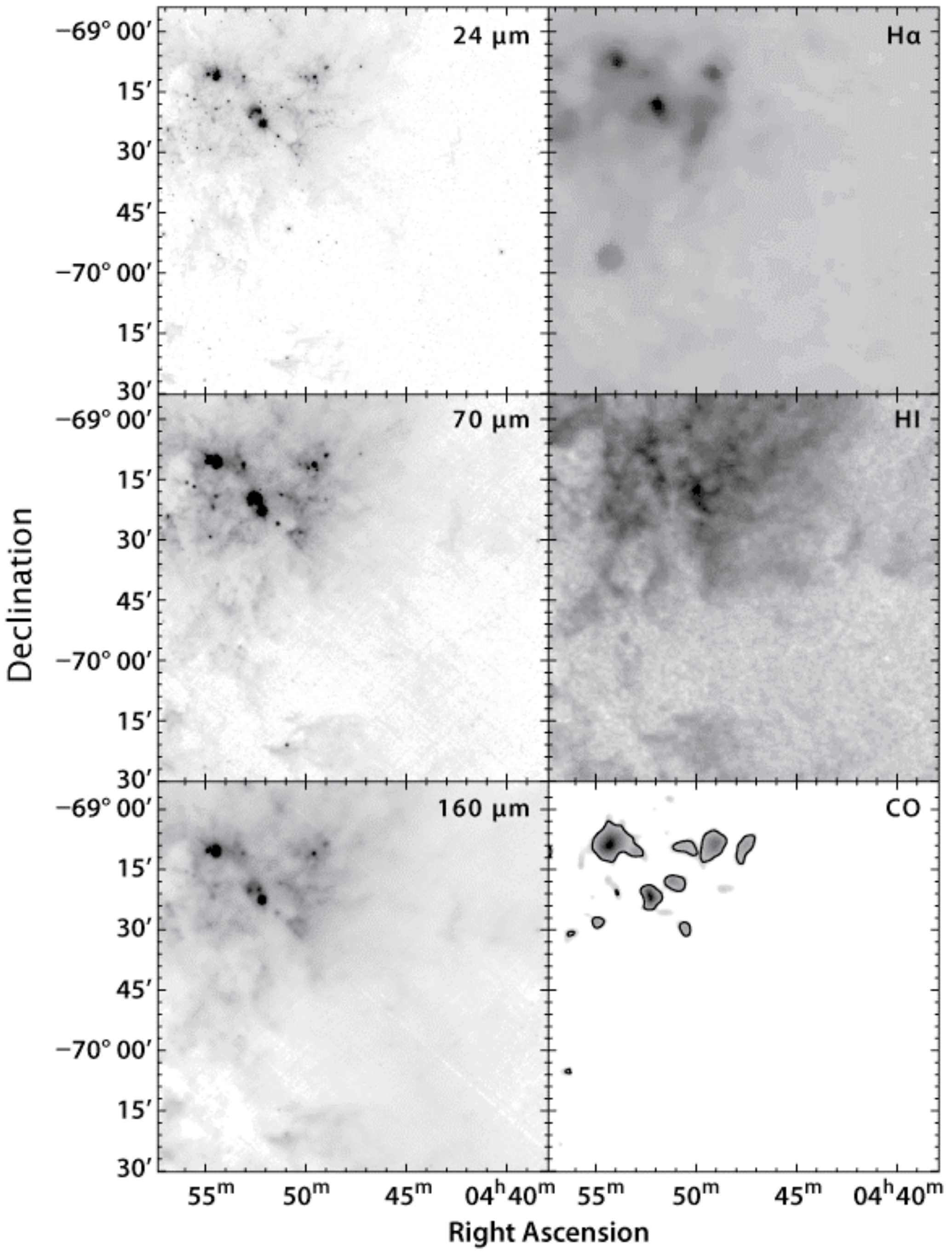

**Figure 9** The MIPS epoch 1 SAGE data of the N79/N83 region in comparison with ISM gas tracers. *Left column*, *Top:* MIPS 24 μm with a square-root greyscale ranging from 0 to 140 MJy/sr*, Middle:* MIPS 70 μm with square-root greyscale ranging from 0 to 130 MJy/sr *Bottom:* MIPS 160 μm with a square-root greyscale ranging from 0 to 350 MJy/sr . *Right column*, *Top:* the Hα SHASSA data (Gaustad et al. 2001), *Middle:* the HI 21 cm line data (Staveley-Smith et al. 2003; Kim et al. 2003), and *Bottom:* the CO J=1-0 line data where the contour marks the 3σ level and the data are absent from the bottom right (southwest) part of the image (Fukui et al. 2001; Fukui et al. 2006).

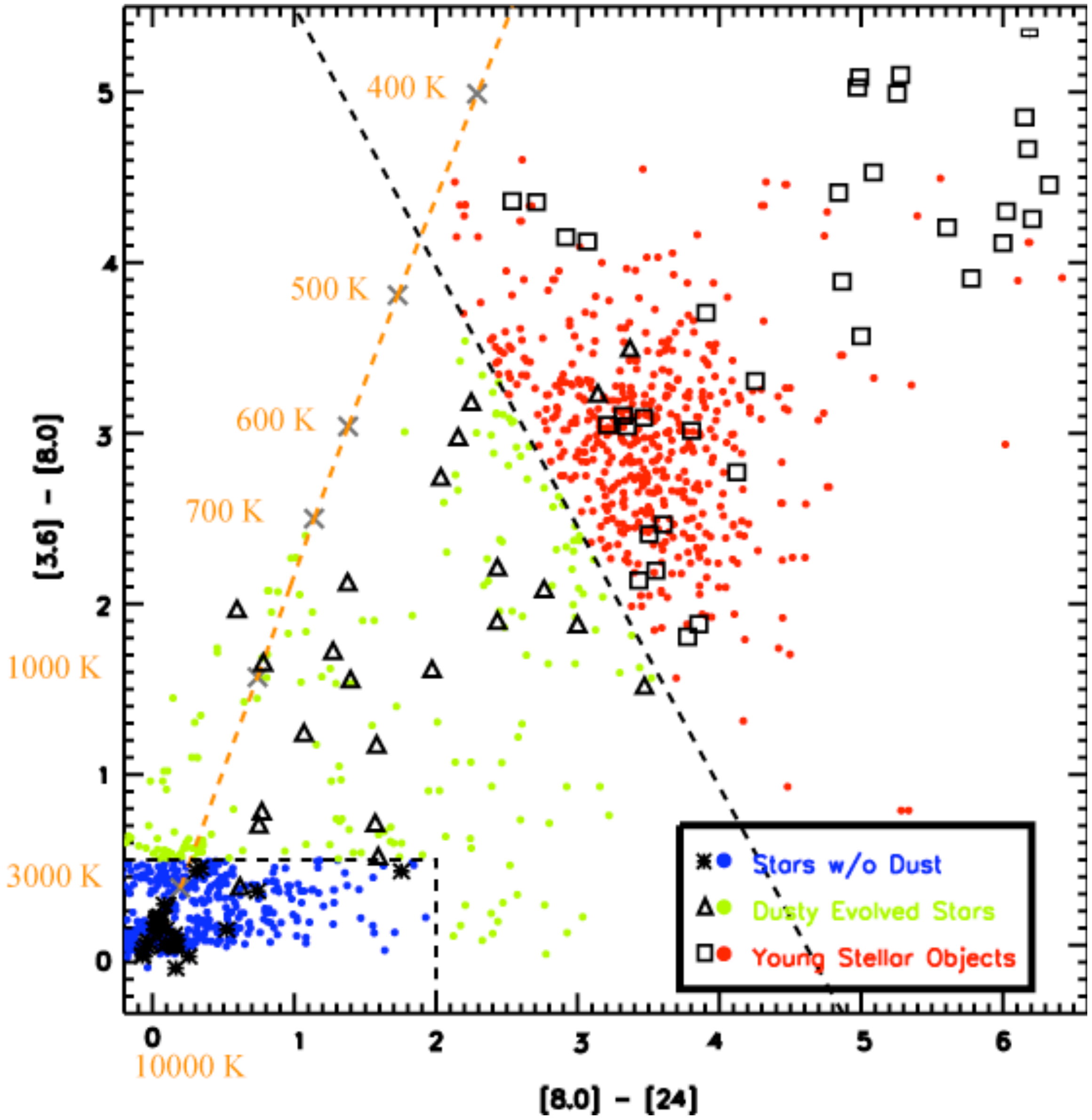

**Figure 10** The [3.6]-[8.0] vs. [8.0]-[24] color-color diagram shows the broadest separation of sources while still retaining enough sources to derive a source classification. For all bands, [ ] means the Vega magnitude at that wavelength, e.g. [8.0] is the Vega magnitude at 8.0 μm. Black symbols are MW templates from Cohen et al. (1993) and Whitney et al. (2004). Asterisks are stars without dust, such as main sequence stars, or red giants. Open triangles are dusty evolved stars. Open squares are young stellar objects. The black dashed lines mark the boundaries for the source classification based on the location of the MW templates. The orange dashed line shows the color temperature for blackbodies with temperatures ranging from 10,000 K to 400 K, marked as grey crosses on the line. The 1175 colored dots are point sources detected in the IRAC 3.6, and 8.0 μm and the MIPS 24 μm from the N79/N83 region. The red, green and blue dots are candidate young stellar objects, dusty evolved stars and stars without dust, respectively. These color classified sources are plotted in the color-color and CMDs of Figure 9 and 10, respectively, to provide guidance on the types of sources plotted.

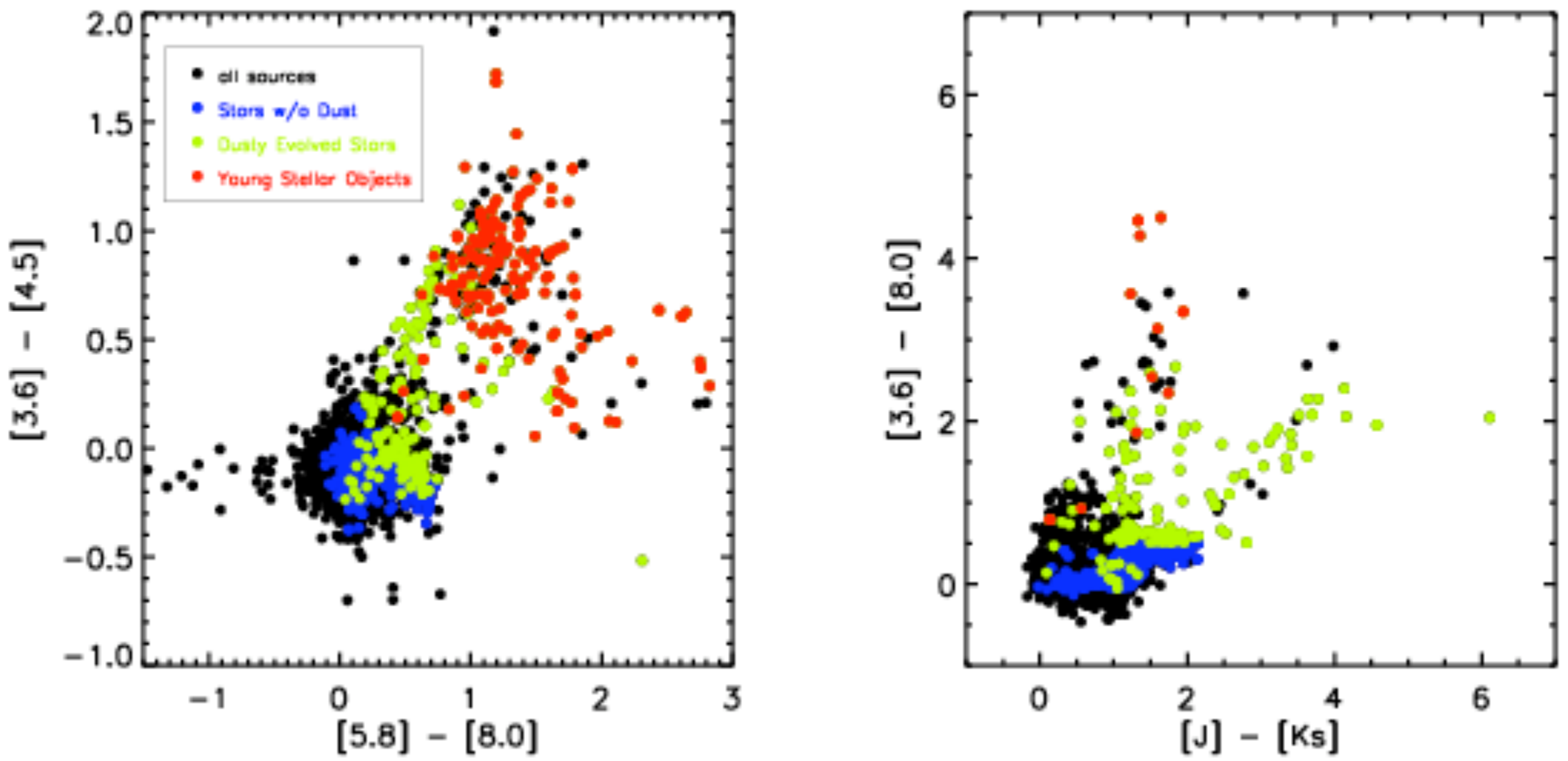

**Figure 11** Color-color plots, *Left:* the IRAC only and *Right:* the IRAC and 2MASS, showing how the source classification holds at other wavelengths for point sources detected in the SAGE N79/N83 region. The black dots are plotted first and include all the sources detected in bands plotted. The colored dots are overlaid on the black dots and include the same dots as plotted in the source classification template (Figure 8).

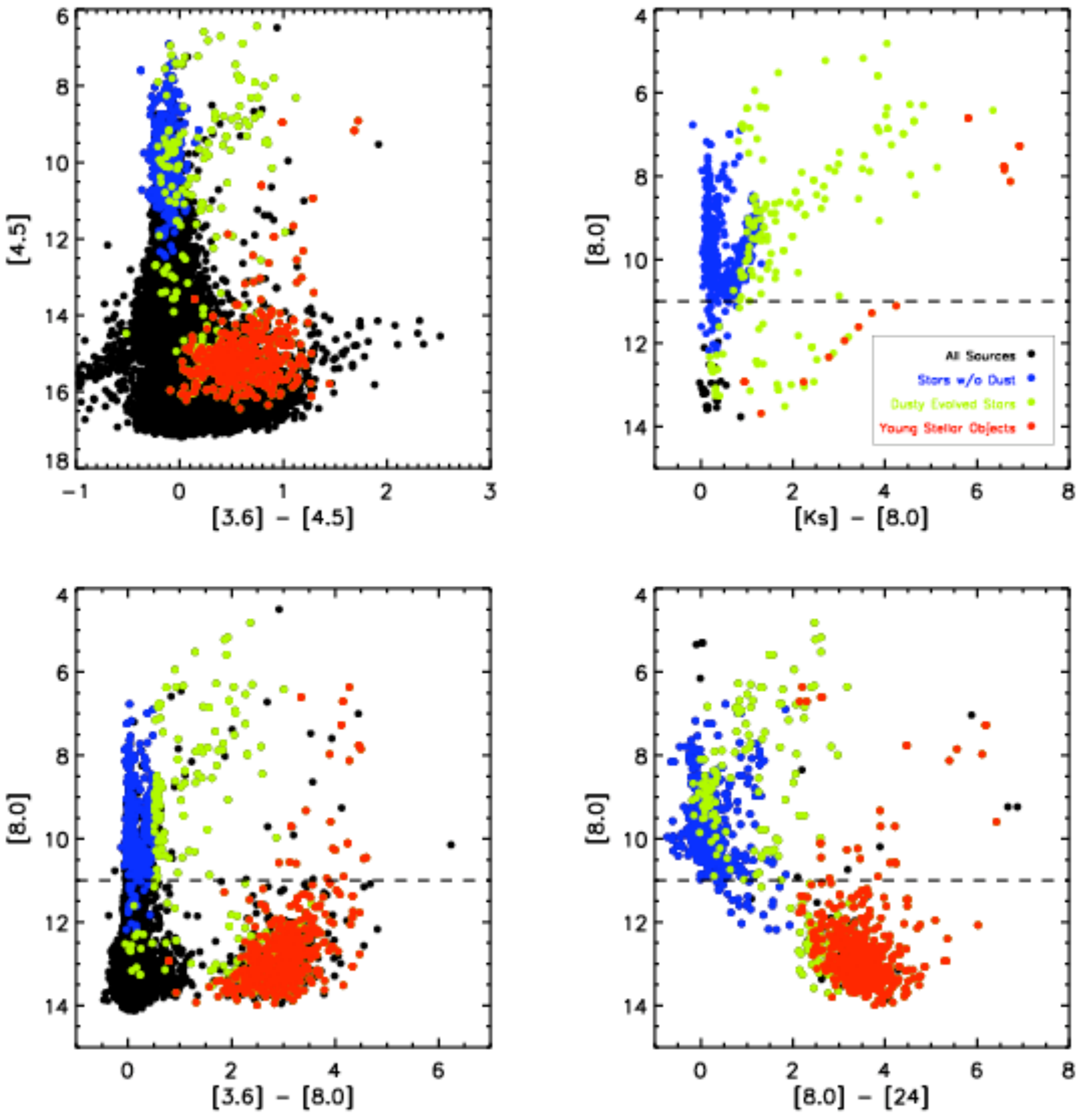

**Figure 12** Four CMDs of SAGE data showing all the point sources from the N79/N83 region in black which have been detected at the wavelengths plotted in the CMD. The candidate classified sources from the source classification are overlaid in the same color used in Figure 8. The black points are sources detected in the bands of the CMD but not in all of the bands of the source classification diagram and hence tend to be fainter sources not detected at the longer wavelengths, e.g. 24 μm. The black dashed line at [8.0] = 11 magnitudes marks the mass-loss rate of $10^{-8}$ $M_{\odot}$ $yr^{-1}$, above which the AGB stars appear to have dusty winds.

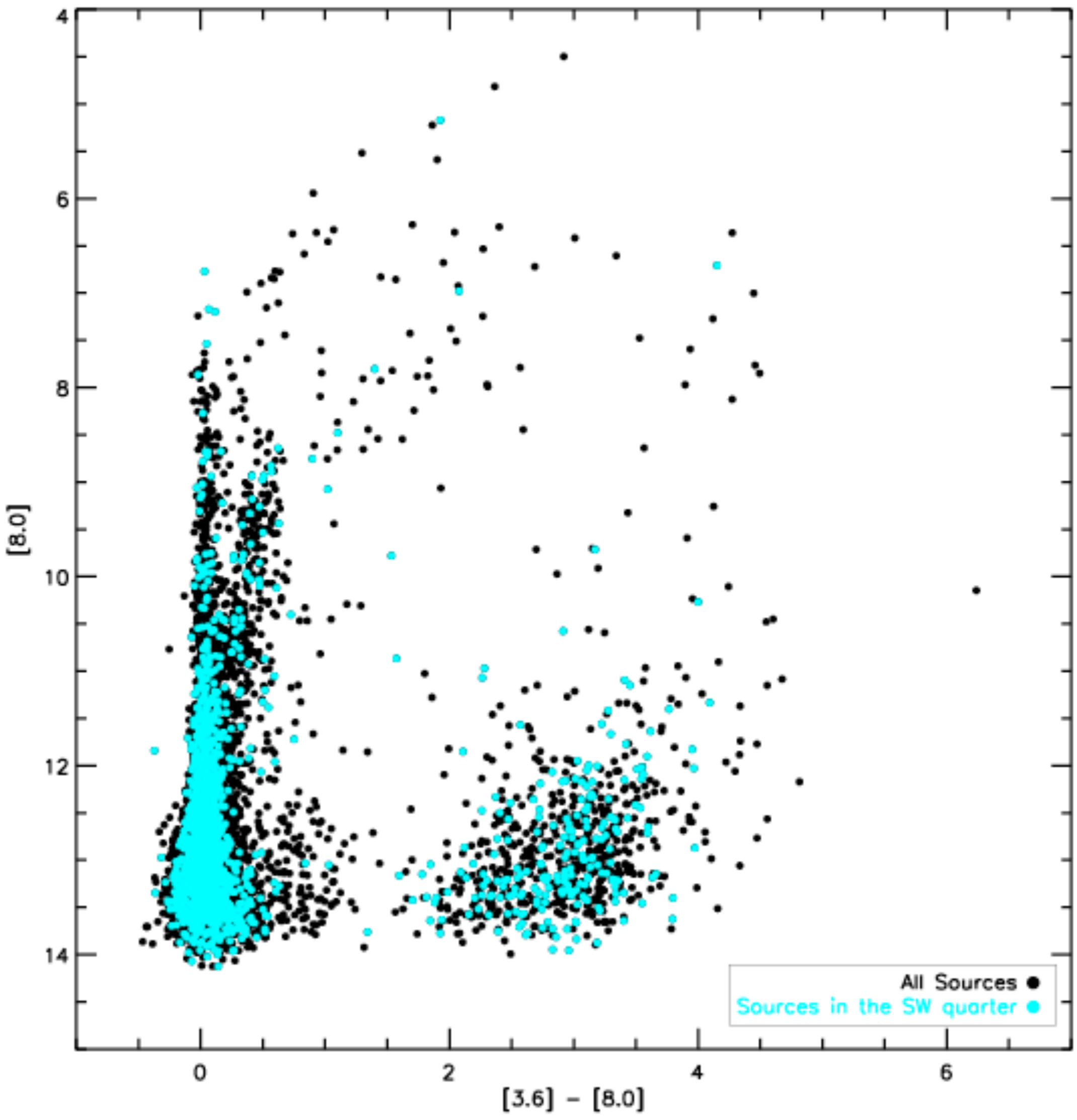

**Figure 13** The [8] vs. [3.6]-[8] CMD for all sources detected in the N79/N83 region at these two wavelengths. Overplotted in cyan filled circles is the subset of sources in the south west quarter of the N79/N83 region that is outside of the HII regions.

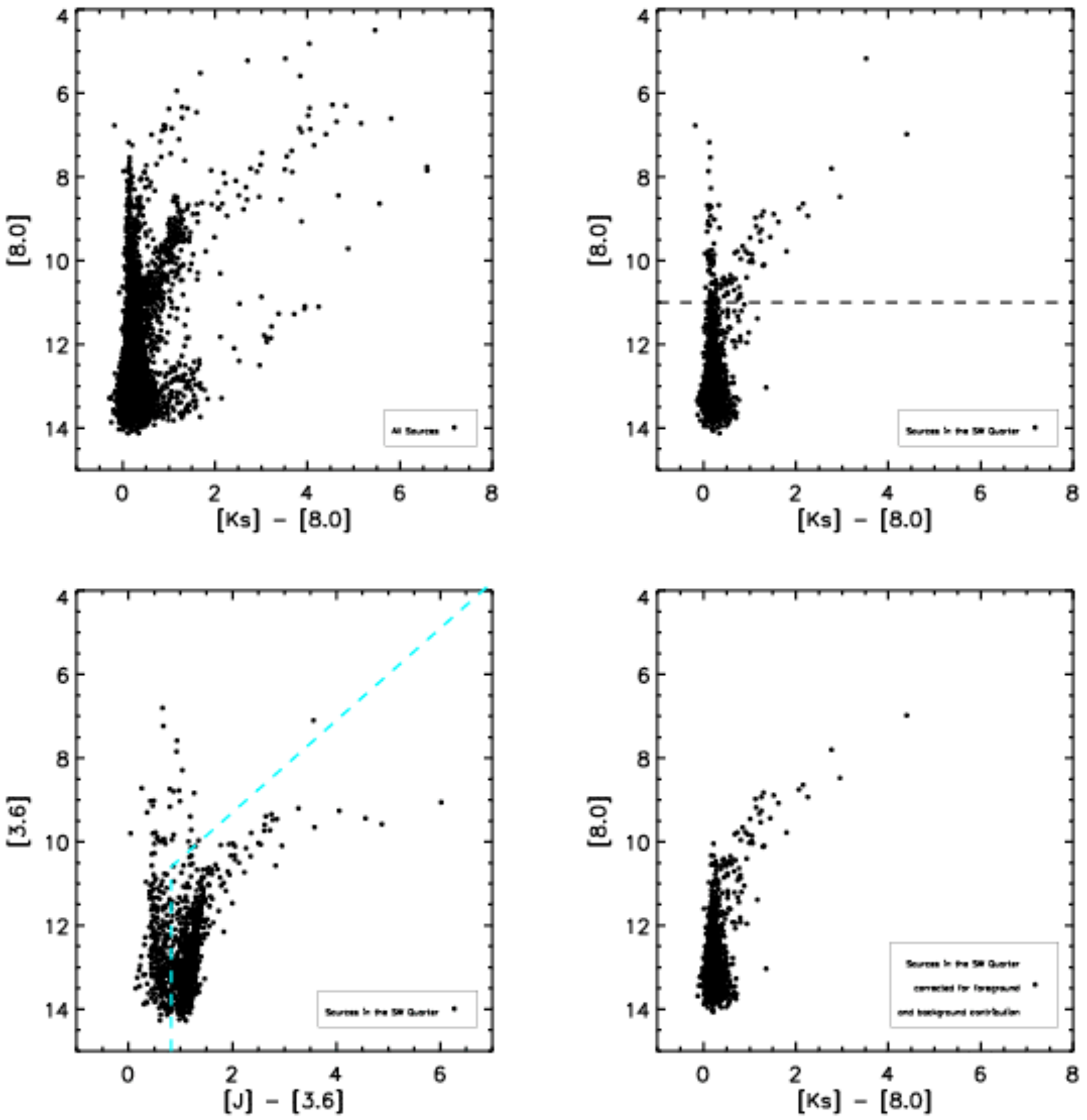

**Figure 14** Near infrared (2MASS) and IRAC CMDs for sources detected in all seven bands in the N79/N83 region (*upper left* panel). The "sw" region (*upper right* panel) contains sources detected in the N79/N83 southwest quarter and has been corrected for background galaxies based on the cuts of Figure 11, see text. The [3.6] vs. J-[3.6] diagram (*lower left* panel) is effective in discriminating between foreground MW objects and LMC giants and AGB stars which should generally lie below and to the right of the *dashed* line (see text). The "corrected sw" diagram shows the remaining sources, primarily LMC AGB stars, after removing foreground MW stars and background galaxy contributions.